\RequirePackage[2020-02-02]{latexrelease}
\documentclass[prd,aps,amssymb,amsmath,superscriptaddress,tightenlines,nofootinbib]{revtex4}
\usepackage{graphicx,epsfig}
\usepackage{slashed}
\usepackage[colorinlistoftodos]{todonotes}
\usepackage{appendix}

\begin{document}

\leftline{KCL-PH-TH/2022-{\bf 36}}
\vspace{1cm}

\title{Primordial Black Holes and Gravitational Waves in Multi-Axion-Chern-Simons Inflation}

\author{Nick E. Mavromatos}

\affiliation{Department of Physics, School of Applied Mathematical and Physical Sciences, National Technical University of Athens, 9 Iroon Polytechniou Str., Zografou Campus, 15780 Athens, Greece}
\affiliation{Theoretical Particle Physics and Cosmology Group, Department of Physics, King's~College~London, Strand, London WC2R 2LS, UK}

\author{Vassilis C. Spanos}

\affiliation{National and Kapodistrian University of Athens, Department of Physics, Section of Nuclear and
Particle Physics, GR-157 84 Athens, Greece}

\author{Ioanna D. Stamou}

\affiliation{National and Kapodistrian University of Athens, Department of Physics, Section of Nuclear and
Particle Physics, GR-157 84 Athens, Greece}

\affiliation{{\rm Address after July 1st 2022}:  Service de Physique Th{\'e}orique, Universit\'e Libre de Bruxelles,
Boulevard du Triomphe CP225, B-1050 Brussels, Belgium} 

\begin{abstract}

\vspace{0.2cm}

We study aspects of inflation and the possibility of enhanced production of primordial black holes (PBHs) and gravitational waves  (GWs) in a string-inspired model of  two axion fields coupled to Chern-Simons gravity, which results in a running-vacuum-model inflation. Fluctuations of the scale invariant spectrum, consistent with the cosmological data, are provided 
in this model by world-sheet (non-perturbative) instanton terms of the axion field arising from string compactification. As a result of such modulations, there is an enhanced production of PBHs and GWs in such cosmologies, which may lead to observable in principle patterns in the profile of GWs during the radiation era. Moreover, we demonstrate that the PBHs may provide 
a significant amount of Dark Matter in this Universe. For comparison, we also discuss a two-stage inflation cosmological model of 
conventional string-inspired axion monodromy, involving again two axion fields. The resulting modifications imprinted on the GWs spectra between these two classes of models 
are distinct, and can, in principle, be distinguished by future interferometers.  
We consider models with more or less instantaneous reheating. We also make some remarks on the effects of a prolonged reheating period in leading to further enhancement of the power spectrum and thus fractions of PBHs that play the r\^ole of Dark matter. 
\end{abstract}

\maketitle

\section{Introduction}

The observation of gravitational waves (GWs) from black-hole mergers by the LIGO-VIRGO collaboration~\cite{LIGO} opened up a new and exciting observational window into the early Universe, which can provide information on constraints on new physics models, that is  complimentary  to the plethora of the other existing precision cosmological data~\cite{Planck,Planck2018}. 
In addition, the recently observed tensions in the current era cosmological data~\cite{tens1,tens2}, mainly the $H_0$ tension on the 
value of the current-era Hubble parameter, but to some extent also the growth of data tension $\sigma_8$, sparked interest among theorists to search for new cosmological models which go beyond the $\Lambda$CDM paradigm~\cite{tens3} and could alleviate such tensions, provided the latter do not admit astrophysical or statistical explanations~\cite{mund}.
One such cosmological framework, which also provides resolutions to both $H_0$ and $\sigma_8$ tension~\cite{rvmtens}, is the so-called Running-Vacuum-Model (RVM)~\cite{sola}, which provides a smooth evolution of the Universe, from a dynamical inflation era, without external inflaton fields, to the current era~\cite{lima}, accounting also for its thermodynamical aspects~\cite{therm}. The model is compatible with the current-era phenomenology,  but also claims in-principle-observable deviations from the $\Lambda$CDM paradigm~\cite{rvmpheno,Tsiapi}. 

In \cite{ms}, a scenario for a cosmological running vacuum model, termed stringy RVM, 
has been presented, based on a (3+1)-dimensional axion-Chern-Simons gravity~\cite{pi,alyunes} obtained 
in the low-energy limit of string theory~\cite{strings}, where the axion is the string-model independent axion~\cite{svrcek,kaloper,kaloper2} that exists in  
the fundamental massless string multiplet, and the Chern-Simons terms are due to the Green-Schwarz anomaly cancellation~\cite{gs}. The model is characterised by a slow-roll {\it linear} axion potential, as a result of primordial GWs condensation, which in turn leads to gravitational-anomaly condensation and inflation of RVM type. The latter is due to the gravitational non-linearities of the vacuum, and not to any external inflaton fields. 
From the point of view of an effective cosmological field theory, the  
model resembles~\cite{mavtorsion} the linear-axion monodromy inflation models arising in some brane-compactification scenarios, characterised by a linear axion potential~\cite{silver}. In those models, however, inflation is induced as a consequence of the linear axion potential, in contrast to our case, in which, as already mentioned, inflation is primarily the result of the RVM form of the vacuum energy density, with the slowly-rolling axion being responsible for the slowly-running features of the induced inflation, compatible with the data~\cite{Planck}. 

Nonetheless, both classes of models are characterised by potentials with linear dependence on the axion fields. We remark at this point, for completeness, 
that a general procedure for understanding and deriving such potentials in field theory 
constructions of monodromy models is provided in \cite{kaloperLin}. 

Non-perturbative stringy effects (world-sheet instantons) can lead to periodic contributions to the effective potentials~\cite{silver}, which can lead in general to enhanced gravitational perturbations during the inflationary period, and enhanced densities of primordial black holes (PBHs). These phenomena may affect the profiles of GWs at the post inflationary (radiation) era, with observable in principle consequences for interferometric studies of the early Universe, and constitute the main topic of discussion in the present work. Moreover, the enhanced production of PBHs during inflation in this model may lead such objects constituting a significant amount of Dark Matter (DM) in the Universe. In the following, we shall restrict most of our attention to effective cosmological theories arising from the model of \cite{ms}, 
a brief review of which is given in the next section. However, for completeness, we shall also compare our results with another class of multi-axion models, inspired from conventional axion-monodromy-induced inflation in string theory~\cite{sasaki}. In such a class of models, there exist a 
two-stage inflation. The compactification axion  drives the first inflationary era, while its world-sheet instanton modulation effects  provide the enhancement of PBHs and GWs produced during inflation.
As we shall discuss, however, the enhancement mechanism is different from that of \cite{sasaki}, since, in our approach here, it is due to a (near) inflection point~\cite{inflectionpoint,inflectionpoint2,Ballesteros:2019hus} the potentials of these models possess naturally.
The duration of inflation in this latter model is prolonged by the KR axion, whose linear potential term (which, in the context of \cite{ms}, could also be due to the aforementioned GWs condensates, that are formed as a result of the enhanced production of GWs) eventually takes over and drives the second inflationary era. We note also that, microscopically, the linear terms pertaining to the compactification-axion that drive the first inflationary epoch might be due to stringy/D-brane compactification effects, which dominate over the gravitational anomaly condensates at the early stages of the inflation. 

We also note, for completeness, that, in agreement with our considerations in this work, although in a different microscopic-model context,  
the r\^ole of non-perturbative-effects-induced modulations of the effective potential of multifield models of axion monodromy in supergravity have been 
discussed recently in \cite{zavala}. It was found that the modulations can enhance the power spectrum to produce a considerable population of light PBHs and large wide modulated GWs spectra, which may be detectable by upcoming GW surveys. Moreover, we mention that a different mechanism for the enhancement of PBHs and GWs during inflation, which does not use the flattening of the potential, but is based on  higher-order modifications of dispersion relations for the scalar perturbations within the context of effective field theories for inflation, is given in \cite{asoor}.

The structure of the article is the following: in the next section \ref{sec:ags} we first review a two-axion-Chern-Simons gravity, which lead to RVM inflation in the approach \cite{ms}. Then we discuss the r\^ole of world-sheet instanton effects that lead to modulations of the effective axion potential. The inclusion 
of such effects induces two cases, depending on the hierarchy of scales involved in the effective two axion potential. In the first case, which is that encountered in the model of \cite{ms}, the KR axion drives the slowly-moving features of the RVM inflation, while in  the second, one encounters the situation of conventional two-axion-monodromy string-inspired models, discussed in \cite{sasaki}. In  section \ref{sec:2mon}, we discuss the enhancement of the power spectrum, and of the densities of PBHs and GWs produced during inflation in the first case, demonstrating observable in principle effects on the profile of GWs in the radiation era, as well as a r\^ole of the PBHs as providers of a significant amount of DM in the Universe. For comparison, in section \ref{sec:sas} we discuss the second case, pertaining essentially to the class of models considered in \cite{sasaki}, but with a different enhancement mechanism, due to the near inflection points the axion potentials have in these models.  In section \ref{sec:reh} we discuss reheating scenarios at the end of inflation that characterise the two classes of string-inspired cosmological models, given that the thermal history of the post -inflationary Universe affect the mass function and hence densities of PBHs that may play the r\^ole of a DM component.   The models we discuss here are characterised by sets of parameters that lead to almost instantaneous reheating, however we also point out that models with prolonged reheating periods will lead to further enhancement of the 
PBHs mass densities, so the coinclusions of this work regarding the enhancement mechanism of cosmic perturbations should be considered as rather conservative. Finally, conclusions and outlook are given in section \ref{sec:concl}. Some technical aspects of our approach, associated with the evaluation of cosmological perturbations in multifield inflation, as well as a review of 
the early stages of cosmological evolution of the RVM, including its reheating aspects at the exit from inflation,  
are given in Appendices \ref{appendix1} and \ref{sec:appRVM}, respectively.

\section{The Axion-gravitational-Chern-Simons Model \label{sec:ags}} 

The string-inspired gravitational model of \cite{ms} deals with effective (3+1)-dimensional cosmological models obtained from string theories after appropriate compactification, with gravitational anomalies. The basic assumption is that only fields from the {\it massless} gravitational multiplet of strings characterise the very early Universe (in the bosonic closed-string sector, these fields are the spin-0 dilaton, $\Phi(x)$, the spin-2 graviton $g_{\mu\nu}(x) =g_{\nu\mu}(x)$, $\mu, \nu=0, \dots,  3$, and the spin-1 antisymmetric tensor $B_{\mu\nu}=-B_{\nu\mu}$.\footnote{In superstring scenarios, the gravitino fields are assumed in ref.~\cite{ms} to acquire heavy masses (close to Planck scale) dynamically, and are thus integrated out of the spectrum of the massless fields, which appear in the effective action of \cite{ms} describing the early eras of the Universe.}) 

The axion, here denoted with $b(x)$, is the so-called string-model-independent axion field~\cite{svrcek}, which in four space-time dimensions is dual to the field strength $\mathcal H_{\mu\nu\rho}$, $\mu, \nu, \rho =0, \dots 3$,  of the antisymmetric (spin one) Kalb-Ramond field, 
\begin{align}\label{Hdef}
\partial_\mu b \propto \varepsilon_{\mu\nu\rho\sigma}\, \mathcal H^{\nu\rho\sigma}~,
\end{align} 
with $\varepsilon_{\mu\nu\rho\sigma} = \sqrt{-g}\, \epsilon_{\mu\nu\rho\sigma}$, where $\epsilon_{0123}= - \epsilon_{1023} = +1$ {\it etc.}, and $\epsilon_{0023} = \epsilon_{0113} =0$ {\it etc.}, the totally antisymmetric Levi-Civita symbol, is the covariant Levi-Civita tensor, with $\varepsilon^{\mu\nu\rho\sigma} = \frac{1}{\sqrt{-g}} \epsilon^{\mu\nu\rho\sigma}$. 

Due to the Green-Schwarz anomaly cancellation  mechanism in string theory~\cite{gs}, the field strength $ \mathcal H_{\mu\nu\rho}$ is modified by the so-called gravitational (``Lorentz'') ($\Omega_{\rm 3L})$ and Yang-Mills (gauge) ($\Omega_{\rm 3Y}$) Chern-Simons (CS) terms (in form language, for notational convenience):
\begin{align}\label{GSH}
\mathbf{{\mathcal H}} &= \mathbf{d} \mathbf{B} + \frac{\alpha^\prime}{8\, \kappa} \, \Big(\Omega_{\rm 3L} - \Omega_{\rm 3Y}\Big),  \nonumber \\
\Omega_{\rm 3L} &= \omega^a_{\,\,c} \wedge \mathbf{d} \omega^c_{\,\,a}
+ \frac{2}{3}  \omega^a_{\,\,c} \wedge  \omega^c_{\,\,d} \wedge \omega^d_{\,\,a},
\quad \Omega_{\rm 3Y} = \mathbf{A} \wedge  \mathbf{d} \mathbf{A} + \mathbf{A} \wedge \mathbf{A} \wedge \mathbf{A},
\end{align}
where $\wedge$ denotes the exterior product among differential ($k,\ell$) forms (${\mathbf f}^{(k)} \wedge {\mathbf g}^{(\ell)} = (-1)^{k\, \ell}\, {\mathbf g}^{(\ell)} \wedge {\mathbf f}^{(k)}$), $\mathbf{A}$ is the Yang-Mills gauge field one form, and $\omega^a_{\,\,b}$  the spin connection one form, with the Latin indices $a,b,c,d$ being  tangent space indices. The 
quantity $\kappa = \sqrt{8\pi \, {\rm G}} = M_{\rm Pl}^{-1}$  (in units of $\hbar=c=1$) is the four-dimensional gravitational constant, with G Newton's constant, and $M_{\rm Pl} =  2.43 \times 10^{18}$~GeV the reduced Planck mass. The  parameter $\alpha^\prime$, on the other hand, is the Regge slope of the string, $\alpha^\prime = M_s^{-2}$, where  $M_s$ is the string mass scale. The latter is in general different to $M_{\rm Pl}$, and in fact it appears to be a free parameter of string theory, to be constrained phenomenologically.   In \cite{ms}, the string scale was constrained under the specific circumstances of the model, which we shall review briefly below.

In the model of \cite{ms}, during the early Universe phase, gauge and other matter fields are assumed absent as external states, and are only generated at the end of the running-vacuum-model (RVM) inflation, which proceeds without the need for external inflation fields, but is due to non-linear gravity effects in the RVM evolution~\cite{sola,lima}. 
Hence, setting the dilaton constant (in a self-consistent way~\cite{bms}), and $\mathbf A=0$ in \eqref{GSH}, we obtain the Bianchi identity 
\begin{align}\label{modbianchi2}
 \varepsilon_{abc}^{\;\;\;\;\;\;\mu}\, \partial_\mu \, {\mathcal H}^{abc} 
 &=  \frac{\alpha^\prime}{32\, \kappa} \,  R_{\mu\nu\rho\sigma}\, \widetilde R^{\mu\nu\rho\sigma}  
\nonumber \\
&  = \frac{1}{\sqrt{-g}}\, \frac{\alpha^\prime}{16\, \kappa} \, \partial_\mu \Big[\epsilon^{\mu\nu\alpha\beta}\, \omega_\nu^{ab}\, \Big(\partial_\alpha \, \omega_{\beta ab} + \frac{2}{3}\, \omega_{\alpha a}^{\,\,\,\,\,\,\,c}\, \omega_{\beta cb}\Big)\Big],
\quad \widetilde R^{\mu\nu\rho\sigma} = \frac{1}{2} \varepsilon^{\mu\nu\alpha\beta}\, R_{\alpha\beta}^{\quad\rho\sigma}\, .
\end{align}
Implementing the identity \eqref{modbianchi2}
in the relevant string-inspired quantum gravity path integral, corresponding to the order $\alpha^\prime$ (perturbative) string-effective action~\cite{strings}, via a pseudoscalar 
Lagrange multiplier field, $b(x)$, and integrating out $\mathcal H_{\mu\nu\rho}$ in the path integral, we obtain the dual Gravitational Axion-Chern-Simons (GACS) theory, described by the action (for constant dilatons)~\cite{svrcek,kaloper,kaloper2}:\footnote{In this work we follow the conventions: $(-, +, +, +)$ for the metric signature, 
$R^\lambda_{\,\,\mu\nu\sigma} = \partial_\nu \Gamma^\lambda_{\, \,\mu\sigma} + \Gamma^\rho_{\,\, \mu\sigma} \, \Gamma^\lambda_{\, \,\rho\nu} - (\nu \leftrightarrow \sigma)$, $\lambda,\mu,\nu,\sigma= 0, \dots 3$
for the Riemann curvature tensor,  with 
$\Gamma^\lambda_{\, \,\mu\sigma} = \Gamma^\lambda_{\, \,\sigma\mu} =
\frac{1}{2}\, g^{\lambda \rho}\,\Big(g_{\rho \mu, \sigma} + g_{\rho\sigma,\mu} - g_{\mu\sigma,\rho} \Big)$
the torsion-free (Riemannian) Christoffel connection, symmetric in its lower indices (the comma denotes ordinary derivative),  $R_{\mu\nu} = R^\lambda_{\, \,\mu\lambda\nu}$ for the Ricci tensor, and $R=g^{\mu\nu} \, R_{\mu\nu}$ for the Ricci scalar.} 
\begin{align}\label{sea4}
S^{\rm eff}_B &=
\; \int d^{4}x\, \sqrt{-g}\Big[ \dfrac{1}{2\kappa^{2}}\, R - \frac{1}{2}\, \partial_\mu b \, \partial^\mu b  -  \sqrt{\frac{2}{3}}\,
\frac{\alpha^\prime}{96 \, \kappa} \, b(x) \, R_{\mu\nu\rho\sigma}\, \widetilde R^{\mu\nu\rho\sigma}
   + \dots \Big]
\end{align}
where the Lagrange multiplier field $b(x)$ became dynamical, and corresponds to the aforementioned string-model independent axion. We note for completion that the classical duality relation \eqref{Hdef} corresponds to saddle points of the constrained under \eqref{modbianchi2} $(b, \mathcal H_{\mu\nu\rho})$ path-integral before the $\mathcal H_{\mu\nu\rho}$ integration.

\subsection{Gravitational-Anomaly Condensates and Running Vacuum Inflation \label{sec:condRVM}}

In \cite{ms} the condensation of primordial gravitational waves (GWs) has been argued to lead to RVM-type inflation without inflation fields. Indeed, such a condensation produces a condensate for the gravitational anomaly ($\mu$ denotes an UV cutoff for the graviton modes~\cite{stephon}\footnote{In \cite{stephon} the condensate has been calculated using Green's functions and classical gravitational wave equations, for the graviton modes. In \cite{Lyth}, the analysis has been performed
by using creation and annihilation operator coefficients for the graviton modes, and in this sense this second method is closer to the spirit of \cite{ms}, that the condensate arises as a result of quantum gravitational fluctuations of space time. There are slight differences in the final expression between these two expressions of the condensates, which however do not affect qualitatively our arguments in this work. For our purposes we adopt the estimates for the anomaly condensate \eqref{rrt2}, which are based on the classical analysis of \cite{stephon}.})
\begin{align}\label{rrt2}
  \langle R_{\mu\nu\rho\sigma}\, \widetilde R^{\mu\nu\rho\sigma} \rangle  &\simeq  
  \frac{16}{a^4} \, \kappa^2\int^\mu \frac{d^3k}{(2\pi)^3} \, \frac{H^2}{2\, k^3} \, k^4 \, \Theta  = \frac{1}{\pi^2} \Big(\frac{H}{M_{\rm Pl}}\Big)^2 \, \mu^4\, \Theta \end{align}
to leading order in the slow-roll parameter 
\begin{align}\label{theta}
 \Theta = \sqrt{\frac{2}{3}}\, \frac{\alpha^\prime \, \kappa}{12} \, H \,  {\dot b} \, \ll \, 1~,
  \end{align}
with $H$ the (approximately) constant Hubble parameter during inflation. The overdot denotes derivative with respect to the cosmic time $t$ in the RW frame. In arriving at \eqref{rrt2}  one uses the isotropy and homogeneity of the RVM Universe, which has been justified by the existence of a first hill-top inflation, preceeding the RVM one,  that characterises the model of \cite{ms}. The slow-roll nature of the $b$ axion field is obtained as a result of  an appropriate solution of the equations of motion for the $b$ field, in the presence of an anomaly condensate, which remains approximately constant during the entire RVM inflationary period, provided the UV cutoff $\mu$ is chosen such that~\cite{ms}:   
\begin{align}\label{muMs}
\frac{\mu}{M_s} \simeq 15 \, \Big(\frac{M_{\rm Pl}}{H}\Big)^{1/2}.
\end{align}
From Planck cosmological data, one has for the Hubble scale during inflation~\cite{Planck}:
\begin{align}\label{PlHI}
 \frac{H}{M_{\rm Pl}} \simeq 10^{-5}.
 \end{align}
 Assuming the validity of the transplanckian conjecture~\cite{ms}, in the sense that there are no graviton modes with momenta higher than the Planck scale, we may naturally take $\mu \sim M_{\rm Pl}$, in which case \eqref{muMs} implies
\begin{align}\label{Msmu}
\mu \simeq M_{\rm Pl} \simeq 4.7 \times 10^3 \, M_s.
\end{align}
 
Parametrising then (in our metric signature conventions, which are opposite of those of ref. \cite{ms})
\begin{align}\label{slowrollb} 
\dot b = - \sqrt{2\epsilon}\, H\, M_{\rm Pl},
\end{align}
with $\epsilon$ a constant slow-roll parameter, of order $10^{-2}$ to match the Planck data~\cite{Planck}, and using \eqref{Msmu}, we then obtain for the condensate \eqref{rrt2} the estimate:
\begin{align}\label{cond}
 \langle R_{\mu\nu\rho\sigma}\, \widetilde R^{\mu\nu\rho\sigma} \rangle \equiv \widetilde \Lambda_0^4 &\simeq \frac{1}{\pi^2}\, \sqrt{2\epsilon} \,
 \frac{\mu^4}{M_s^2\, M_{\rm Pl}^2}\, H^4 \simeq 3.16 \times 10^5 \, H^4 \,\stackrel{Eq.\eqref{PlHI}}{\simeq} 3.16 \times 10^{-15} \, M_{\rm Pl}^4    \, \nonumber \\ &\Rightarrow \, \widetilde  \Lambda_0 \simeq 2.37 \times 10^{-4}\, M_{\rm Pl}.
 \end{align}
As follows from the form of the effective action \eqref{sea4}, 
the presence of the condensate \eqref{cond} leads to an effective potential for the axion field $b$ of the form:
\begin{align}\label{linV}
&V(b) \simeq b \, \widetilde \Lambda_0^4 \sqrt{\frac{2}{3}} \, \frac{M_{\rm Pl}}{96 \, M_s^2} \equiv b \, \frac{\widetilde \Lambda_0^4}{f_b} \, \equiv b \, \Lambda_0^3 \, \stackrel{Eq.\eqref{Msmu}}{\simeq} \, b \Big( 5.9 \times 10^{-10}\Big) \, M_{\rm Pl}^3, \nonumber \\
 &{\rm with} \quad f_b \equiv  \Big(\sqrt{\frac{2}{3}}\, \frac{M_{\rm Pl}}{96 \, M_s^2} \Big)^{-1} \stackrel{Eq. \eqref{Msmu}}{\simeq} 5.3 \times 10^{-6} \, M_{\rm Pl} \quad {\rm and} \quad 
\Lambda_0  = 8.4 \times 10^{-4} \, M_{\rm Pl}.
\end{align}
We stress once more at this point that the result of the condensate appears to lead~\cite{mavtorsion} to an effective action similar to the one appearing in the so-called axion monodromy stringy linear inflation~\cite{silver}, in the sense that the linear potential for the axion field leads in general to inflation. However in the scenario of \cite{ms}, this linear potential is {\it not} the one that drives inflation. From the form of the condensate \eqref{cond}, which depends on the fourth power of the (approximately constant) Hubble  parameter, we deduced in \cite{ms} that it is these linearities that drive the inflation, which is of the RVM type~\cite{lima,sola}, rather than the axion itself $b$ or any external inflaton field. The field $b$, as already mentioned simply implies a slow-roll evolution as a result of its equations of motion stemming from \eqref{sea4}, in the presence of the condensate. 

We note, though, at this point that the
approximate constancy (slow-roll)  of the term $b \Lambda_0^4$ during the entire inflationary era
requires~\cite{ms} transplanckian values for the (boundary) value of the field $b$ at the beginning of inflation, which may be in conflict 
with the swampland distance conjecture~\cite{swamp}, which is known to afflict the slow-roll field theory models of inflation~\cite{swampinfl}. This conjecture, however, may not be actually realised in all string models. In the context of non-critical string theory models of inflation, for instance, 
which the time-varying $\Lambda(t)$ stringy RVM model of \cite{ms} may be embedded to~\cite{emnlt},
the work of \cite{emnswamp} has argued that the distance and swampland conjectures in the deep quantum gravity regime of the cosmological model may be evaded, consistently with the validity of the slow-roll assumptions on inflation in such models. On the other hand, for the type of string-inspired models studied in \cite{silver}, with inflation driven primarily by linear axion monodromy potentials, the analysis of \cite{heisen} has argued that the incorporation of towers of string states, as required by the distance conjecture~\cite{swamp}, due to the transplanckian values of the axion-inflaton in these models,  can lead, under some circumstances, to trapped inflation, in a way
consistent with the swampland criteria. This feature arises as a consequence of the production of string moduli fields, assumed strongly coupled to the axion-inflaton, 
at enhanced symmetry points at which they become massless. We shall not delve further into such a discussion in our approach here.  In the current work, we shall discuss enhanced GWs and PBHs production during inflation in the stringy RVM of \cite{ms}, upon taking into account the effects of stringy world-sheet instantons. The latter lead to important modulations of the axion-potential, which we proceed to study in the next subsection.

\subsection{Effects of world-sheet instantons in multi-axion string-inspired effective theories \label{sec:wsinst}}

In the presence of world-sheet instantons (wsinst), which are compatible with the basic assumptions of the GACS model of \cite{ms} that only fields from the massless gravitational string multiplet appear as external fields in the early Universe eras, one may obtain periodic potentials for the axion fields, of the form
\begin{align}\label{fbval}
V^b_{\rm ws inst} \simeq \Lambda_b^4 \, {\rm cos}\Big(\frac{b}{f_b}\Big),
\end{align} 
with $f_b$ defined in \eqref{linV}. 

The scale $\Lambda_b \ll \Lambda_0$, since the scale $\Lambda_b$ is suppressed by an exponential of the large (Euclidean) world-sheet or world-volume instanton action $S_{\rm wsinst} \gg 1$, 
\begin{align}\label{wsinst}
\Lambda_b^4 \sim M_s^4 \, e^{-S_{\rm wsinst}}\,.
\end{align} 
Hence, 
\begin{align}
\Lambda_b \ll \Lambda_0\,. 
\end{align}
In the GACS model~\cite{ms}, therefore, such  effects are subleading compared to the anomaly condensate which 
drives the RVM  inflation. 

However, in string models one has another kind of axions, $\widetilde a_I$, $I=1, \dots N$, the ones arising from compactification.
In the context of the GACS model~\cite{ms}, such axions will also couple to the gravitational Chern Simons terms, but their axion constants $f_{a_I}$ will depend on the details of compactification, and as such they will be different from $f_b$. One may conjecture on the existence of appropriate compactifications leading to some of the axions $\widetilde a_I$, $I=1, \dots M < N$, having constants  $f_{a_I} \ne f_b$. 
As the kinetic terms of these axions will come in the form (after appropriate diagonalisation of the metric in field space~\cite{svrcek,silver}) $f_{a_I}^2 \partial_\mu \widetilde a_I \partial^\mu \widetilde a_I $, canonical normalization of the axions 
\begin{align}\label{anorm}
f_{a_I}\, \widetilde a_I \equiv a_I\,,
\end{align} 
will lead to coupling to anomalous terms multiplied by large $f_{a_I}^{-1}$ factors.

In the presence of the condensate \eqref{cond}, such axions will also acquire linear potential (density) terms, of the form 
\begin{align}\label{linpota}
V^{a_I}_{\rm lin} = a_I(x)\,\frac{f_b}{f_a} \Lambda_0^3 \,. 
\end{align}
World-sheet instanton effects will also generate periodic potentials 
\begin{align}\label{wsinsta}
V_{\rm wsinst}^{a_I} \simeq \Lambda_I^4 \, {\rm cos}\Big(\frac{a_I}{f_{a_I}}\Big)\,, 
\end{align}
with $\Lambda_0 \gg \Lambda_I \ne \Lambda_b $, in general. This ensures that the basic assumption of \cite{ms}, according to which 
RVM inflation is driven by the gravitational-anomaly condensate term for the KR axion $b$, is satisfied.

For our considerations below we consider for simplicity the concrete case where only one stringy axion, arising from compactification, plays a r\^ole during the early epoch of the Universe in the model of \cite{ms}, in the presence of the condensate \eqref{cond},
 \begin{align}\label{I1}
I=1 : \,\, a_1 \equiv a \,.
\end{align}

We now remark that, in brane compactifications of, say, type IIB string theories~\cite{silver},  {\it e.g.} D5 or NS5 branes wrapped around appropriate two cycles, one obtains, as  a consequence of the so-called DBI world-volume action on the brane, additional linear contributions to the potential of the axion $a$, and also
modulations of the world-sheet instanton scale $\Lambda_1$ by terms linear in $a$, for large field values $a \gg M_s$, appropriate for inflation: 
\begin{align}\label{compacteffects}
V_{brane-compact.-effects} (a) \ni  \Lambda_2^4 \, \frac{1}{f_{a}} \, a  + 
\Lambda_1^4 \Big(1 + \xi_a \, \frac{a}{f_{a}} \Big) \, {\rm cos}\Big(\frac{a}{f_{a}}\Big).  
\end{align}

In the type IIB string model of \cite{silver}, involving D5-branes, wrapped around a two cycle $\Sigma^{(2)}$ of size 
$\ell \sqrt{\alpha^\prime} = \ell \, M_s^{-1}$, when compactification (dimensionless) axions $b$ are turned on, 
 the corresponding linear axion potential assumes the generic form 
\begin{align}\label{L1silver}
V_{\rm D5} \sim \frac{\epsilon}{g_s \, (2\pi)^5}\, \sqrt{\ell^4 + b^2}\, M_s^4  \, ,
\end{align}
where $g_s$ is the string scale, and $\epsilon$ is a parameter depending on the appropriate warp factor characterising the detailed string models under consideration~\cite{silver}. The dimensionless axions $b$ are related to the canonically normalised axion $a$ as~\cite{silver}: 
\begin{align}\label{typeIIB}
a^2 \sim \frac{L^2}{3g_s^2 \, (2\pi)^7}\, b^2\, M_s^2 
\sim \frac{1}{6\, L^4} \, b^2 M_{\rm Pl}^2 \,,
\end{align}
in simplified scenarios of compactification of the extra six-dimensional manifold with one compactification-radius-scale involved, $L\sqrt{\alpha^\prime}$, such that the compactified  volume is $\mathcal V^{(6)} \sim  (L \, \sqrt{\alpha^\prime})^6$. In arriving at the second equality  of \eqref{typeIIB} 
we took into account the type IIB  basic string compactification relation~\cite{strings,silver}: $\alpha^\prime M_{\rm Pl}^2 \sim \frac{2\, \mathcal V^{(6)}}{(2\pi)^7\, g_s^2}$. 

For large $b \gg \ell^2 $ axions, as required by the consistency of the inflationary scenario of \cite{silver} with cosmological data~\cite{Planck}, one obtains from \eqref{L1silver} a linear axion-$a$ potential term of the form \eqref{compacteffects}, corresponding to a scale 
\begin{align}\label{L2} 
\frac{\Lambda_2^4}{f_a}  \sim \frac{\epsilon}{L}\, \sqrt{\frac{3}{(2\pi)^3}}\, M_s^3\,,
\end{align}
with $\epsilon$ depending on the details of the appropriate brane warp factors, as mentioned above.

On the other hand, for the specific type IIB brane-compactification models of \cite{silver}, 
the $a \, {\rm cos}(a/f_{a})$ structures in \eqref{compacteffects}  have been assumed sufficiently small, so as not to affect the slow-roll inflation driven by the linear potential.  The consistency of the axion-monodromy inflation of the model of \cite{silver} with the data, requires large-$a$ field inflation, taking on transplanckian values at the beginning of the inflationary eras, which are in conflict with the swampland distance conjecture~\cite{swamp}, as mentioned previously.
In the context of the model of \cite{ms}, when adapted to contain such brane compactifications, a similar assumption can be made, so that the main features of the GW-induced-condensate-\eqref{cond} driven RVM inflation remain 
unaffected. However in our work we shall keep the modulated amplitude feature of the amplitude of the periodic (sinusoidal) 
world-sheet-instanton corrections to the effective potential of the axions $a_I$ arising from compactification. This may lead to in principle observable effects on the features of GWs during inflation, as we shall discuss in the present article.

Thus, in this generalised context, the corresponding world-sheet instanton terms may not be negligible, depending on details of compactification. Indeed, 
in view of \eqref{anorm}, we observe that, for sufficiently small
\begin{align}\label{condfa}
\Big(\frac{f_b}{f_a}\Big)^{1/3} \, < \, \frac{\Lambda_1}{\Lambda_0}\, \,
\stackrel{\rm Eq. \eqref{linV}, Eq.\eqref{wsinst}}{\simeq} \,\, e^{-\frac{1}{4}\,S_{\rm wsinst}}\, 10^{-3} \, \frac{M_s}{M_{\rm Pl}} 
\stackrel{\rm Eq. \eqref{Msmu}}{\simeq} 2 \, e^{-\frac{1}{4}\,S_{\rm wsinst}}\,\times 10^{-7} \,\,
\,,
\end{align} 
where, without loss of generality, we assumed that, in order of magnitude, one has $\Lambda_1 \sim \Lambda_b$ (see \eqref{wsinst}), 
the scale of the world-sheet instanton potential term \eqref{wsinsta} for the (compactification-induced) $a$ axion becomes larger than the corresponding scale of the linear term \eqref{linpota} for this axion. As we shall show in the next section, this suffices to induce enhancement in the density of primordial black holes and gravitational-wave perturbations during the RVM inflation, induced by the linear condensate $b$-term \eqref{linV}.

We now note that the world-sheet-instanton-generated axion periodic potential terms, will generate masses for the axions.
The hierarchy of the induced axion masses depend on the relative strength of $f_b/f_a$. From the quadratic-in-the-field terms in the expansion of the cosines in \eqref{wsinst}, \eqref{wsinsta}, we obtain, on assuming $\Lambda_b \sim \Lambda_1$:
\begin{align}\label{aximass}
\frac{m_a}{m_b} &\sim \frac{f_b}{f_a} \, \stackrel{\rm Eq.\eqref{condfa}}{<} \, 8  \, e^{-\frac{3}{4}\,S_{\rm wsinst}}\,\times 10^{-21}\,, \quad {\rm with} \quad m_b \sim 10^{-2}\, e^{-\frac{1}{2}\,S_{\rm wsinst}}\, M_{\rm Pl} \nonumber \\
\Rightarrow \quad m_ a &< 8 \, e^{-\frac{5}{4}\,S_{\rm wsinst}} \, \times 10^{-23}\, M_{\rm Pl} \,.
\end{align}
For world-sheet instanton actions of $\mathcal O(40)$ this yields ultralight axion with masses $m_a < 1.5 \times 10^{-44} \, M_{\rm Pl} \sim 3.6 \times 10^{-17}$~eV. We remark that such a range of axion masses can still satisfy the requirement that axions obtained from string compactifications can play the r\^ole of (dominant species of) Dark Matter (DM) in the Universe~\cite{marsh}.  

We next remark that, if $\Lambda_b \sim \Lambda_1$, then both periodic contributions from the axions $b$ and $a$, \eqref{fbval} and \eqref{wsinsta}, respectively, appear on equal footing in the effective potential. However, if   
we want to stay closer to the spirit of \cite{ms}, we may consider models in which  $\Lambda_1 \gg \Lambda_b$. In such a case, as we shall discuss below, it is the compactitication-axion world-sheet instanton effects that constitute the dominant sources of fluctuations in the scale invariant spectrum of the RVM inflation associated with the $b$ field. Then, to a good approximation, one may ignore the world-sheet instanton contributions of the $b$ axion, when compared with the rest of the terms in the potential. This is what we assume in what follows.  

Hence, according to the previous discussion, we consider  
effective axion potentials of the form
 \begin{equation}\label{effpot}
 V(a, \, b)={\Lambda_1}^4\left( 1+ f_a^{-1}\, \tilde \xi_1 \, a(x) \right)\, \cos({f_a}^{-1} a(x))+\frac{1}{f_{a}}\Big(f_b \, {\Lambda_0}^3 + \Lambda_2^4 \Big) \, a(x) + {\Lambda_0}^3\, b(x), 
 \end{equation}
 where the parameters $\tilde \xi_1, f_a$, and the world-sheet instanton induced scales $\Lambda_1, \Lambda_2 $ are treated as phenomenological, with the constraint though that 
 \begin{align}\label{constr}
 \Big(\frac{f_b}{f_a} + \frac{\Lambda^4_2}{f_a\, \Lambda^3_0} \Big)^{1/3}\, \Lambda_0  \, < \, \Lambda_1 \ll \Lambda_0 ~, 
 \end{align}
which ensures that the dominant effects in the potential come from the anomaly condensate, so that the spirit of \cite{ms}, regarding the induced RVM inflation, is maintained. For the specific brane-compactification models of \cite{silver} $\Lambda^4_2/f_a$ is given by \eqref{L2}, and we have arranged the compactification parameters in such a way that the hierarchy of scales \eqref{constr} applies, which as we shall discuss 
leads to an enhancement of GWs perturbations and primordial black hole densities, produced during inflation. This might have observable features in interferometers, since the profiles of GWs during the (post-inflationary) radiation era will be affected.

For the model of \cite{ms}, where we restrict our attention here, the value of the constant $f_b$ is given in \eqref{fbval}, and that of the condensate $\Lambda_0$ in \eqref{linV}. Also, to simplify things, we may ignore the term $\Lambda_2^4$ in front of 
$\Lambda^3_0 f_b$ in \eqref{effpot}. The RVM inflation is driven by the last term on the right-hand-side of \eqref{effpot}, linear in the KR axion field $b$, according to the mechanism suggested in \cite{ms}. 

A phenomenological model with axion potentials of the form \eqref{effpot} has been considered in \cite{sasaki} with the purpose of studying features of GWs during inflation, as well as the potential enhancement of PBHs and their potential r\^ole as DM in the Universe~\cite{PBHDM}. The condensate phase in the model of \cite{ms}, makes an apparent microscopic link~\cite{mavtorsion} of such string-inspired models with the 
phenomenological case of \cite{sasaki}, but with important differences as far as the hierarchy of the various scales involved, and the modulation of the periodic instanton-induced terms are concerned, which in our model  assume the form \eqref{effpot}.
In the model of \cite{sasaki} the axion field $a(x)$ dominates first, while the $b$ field contributes to an increased duration of inflation, that is one has the hierarchy:
 \begin{align}\label{constr3}
 \Lambda_0 \, \ll  \, \Big(\frac{f_b}{f_a} + \frac{\Lambda^4_2}{f_a\, \Lambda^3_0} \Big)^{1/3}\, \Lambda_0  \, < \, \Lambda_1 ~, 
 \end{align}
in contrast to our case \eqref{constr} \cite{ms}. 

It is the purpose of this work to discuss the effects of periodic potential modulations on the profile of GWs, as well as the density of PBHs, during the inflationary period, in the context of the model \eqref{effpot}, with the hierarchy of scales \eqref{constr}. 
This  be the topic of discussion in the next section. Comparison of the results with the features of the model of \cite{sasaki} will also be made in section 
\ref{sec:sas}, for completeness. 

\section{Gravitational Waves and Primordial Black Holes  in an axion-monodromy two-field inflation \label{sec:2mon}}

Following our previous discussion, in this section we consider the following Lagrangian density,  in order to obtain an important enhancement of the scalar power spectrum
\begin{equation}
    \mathcal{L}=-\frac{1}{2}(\partial_{\mu}a)^2-\frac{1}{2}(\partial_{\mu}b)^2- V(a,b).
\end{equation}
According to our discussion in the previous section, we consider potentials of the form:
\begin{equation}
     V(a,b)=  g_1   {\Lambda_0}^3a+g_2{\Lambda_0}^4 \cos\left(\frac{a}{f}\right)\left(1+\xi \frac{a}{M_{\rm Pl}}\right)+  {\Lambda_0}^3 b \, ,
\label{eqpot}     
\end{equation}
where, in comparison with \eqref{effpot}, we used the following parametrization, for notational convenience:  
\begin{align}\label{defs}
\quad g_1 \stackrel{\rm Eq.\eqref{constr}}{\equiv} \frac{f_b}{f_a} + \frac{\Lambda^4_2}{f_a\, \Lambda^3_0}  \,, \quad
\Lambda_1^4 \equiv g_2\, \Lambda_0^4, \quad f \equiv f_a\,,  \quad  \xi \equiv  \frac{M_{\rm Pl}}{f_a}\, \tilde \xi_1\,.
\end{align}
Moreover, for notational convenience, we define the scale $\Lambda_3$:
\begin{equation}
\Lambda_3 \equiv (g_1 {\Lambda_0^3})^{1/3}\, .
\end{equation}
If, as mentioned above, in the spirit of \cite{ms}, we ignore the $\Lambda_2$ term (as, either non-existing, due to a different type of compactification compared to that in the work in \cite{silver}, or subleading, compared to the gravitational-anomaly-condensate term~\cite{ms}), then
\begin{align}\label{constr2}
 g_1^{1/3} \simeq \Big(\frac{f_b}{f_a}\Big)^{1/3}\,\, \stackrel{\rm Eq.\eqref{condfa}}{<} \,\,1\, .
\end{align}

In what follows, we shall first study the case \eqref{constr2}, with the scale hierarchy \eqref{constr}, which, using the notation \eqref{defs},
becomes
\begin{align}\label{constrl3}
\Lambda_3  \, < \, \Lambda_1 < \Lambda_0 ~.
 \end{align}
and then, we shall compare the predictions with those of the case \eqref{constr3}, adopted in the model of \cite{sasaki}.

\begin{figure}[h!]
\centering
\includegraphics[width=100mm]{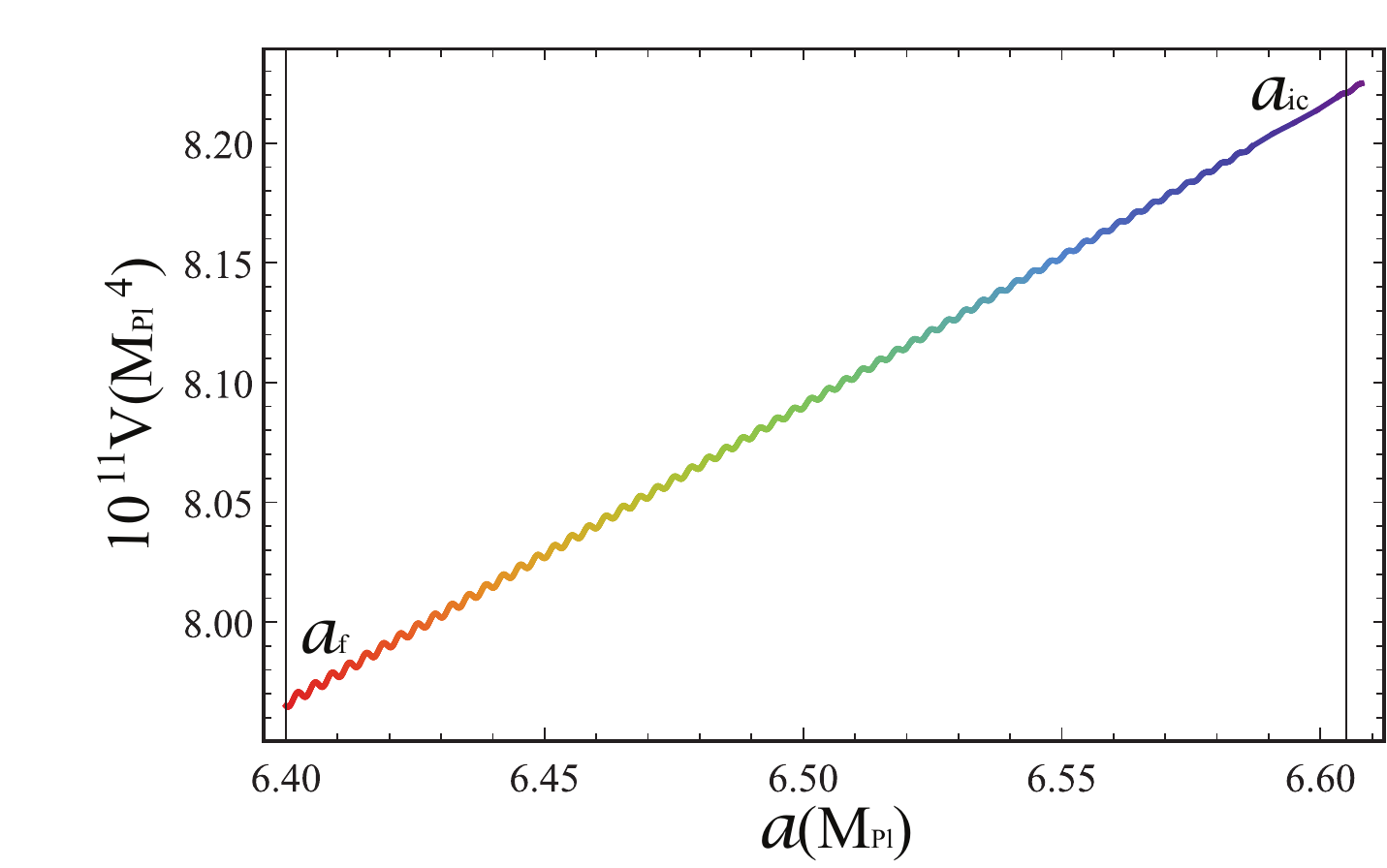}
\caption{\textcolor{black}  The  potential in $a$-field direction for the first set of parameter of Table \ref{table1}. $a_{ic}$ represents the initial condition before oscillations start to be appreciable  and $a_{f}$ the value of the field when the inflation ends. }
\label{f00}
\end{figure}

\subsection{Cosmological Perturbations--enhancement of scalar power spectrum}

In Table~\ref{table1}  we show two sets  of parameters of the potential  \eqref{eqpot}, which can lead to an {\it amplification} in the scalar power spectrum. 
There are many  mechanisms proposed for an enhancement of the  scalar power spectrum  ~\cite{sasaki,Graham:2015cka,Espinosa:2015eda,tetradis,Clesse:2015wea,inflectionpoint,inflectionpoint2}.  The amplification in our analysis comes from the oscillations of the world sheet instantons, associated with the cosine term in \eqref{eqpot}. 
In particular, when the bumps due to the world-sheet-instanton term of Eq.\eqref{eqpot}, proportional to $\Lambda_1^4$, are sufficiently small, the potential  behaves like a linear inflationary potential in the fields plane $(a,b)$. At this stage the dominant term is the $b$ field, as one can notice from the hierarchy in  the Eq.\eqref{constrl3}. This is in the spirit of \cite{bms,ms}, in which framework it is the anomaly condensate that drives the stringy RVM inflation.
 The oscillation pattern  at this time has a negligible effect on the evolution of the fields and this leads to the scale invariance of the  spectrum. This can be seen in Fig.~\ref{f00}, where  in the proximity of the initial condition of the field $a$, $a_{ic}$, the potential in the direction of the $a$ field  is linear.

 At later  times of the Universe the magnitude of the field $b$ takes on smaller values. Hence,  the oscillatory term  proportional to $\Lambda_1^4$  becomes important. The oscillations at this stage are pronounced and constitute the dominant effect, leading to an enhancement of the primordial spectra.  The  values of $a_{ic}$ and $a_f$ presented in Fig.~\ref{f00} denote the initial condition and the value of the $a$-field at the end of inflation, repsectively.  
 They are given  later on in the section, in Table~\ref{table2}.
 The evolution of the fields is determined from the  background equation. 
Specifically, the equations of motion for the fields, when the latter are expressed in terms of the number of e-folds, $N$, are given by the following expression:

\begin{equation}
\label{21}
{ \phi_i''} +3 \phi_i'- \frac{1}{2} \left( \phi_i' \right)^3 +\left(3- \frac{1}{2} \left(\phi_i' \right)^2\right) \frac{d\ln V(\phi_i)}{d \phi_i}=0\, , 
\end{equation}  
where  $\phi_i$ refers to the axion fields $a,b$ and the prime denotes derivative with respect to the e-folds number $N$.  We note that, from now on, we work in Planck units. In this notation, the Friedman equation assumes the form
\begin{equation}
 \label{23}
H^2=\frac{V}{(3- \varepsilon_H)} \, , 
\end{equation}
\noindent
where the slow-roll parameters $\varepsilon_H$ and $\eta_H$ are defined as:
\begin{equation}
 \label{22}
\varepsilon_H=\frac{1}{2}  \sum_{\phi_ i}  \left( \frac{d \phi_i}{dN} \right)^2, \quad \eta_H= \varepsilon_H- \frac{1}{2} \frac{d \ln   \varepsilon_H}{dN} \, ,
\end{equation}
and the reader should recall that inflation stops when $\varepsilon_H > 1$.

\begin{center}
\begin{table}[ht]
\begin{tabular}{||c| c c c c|c c c||} 
 \hline
SET & $g_1$ & $g_2$ & $\xi$ &$f(M_{Pl})$ &$\Lambda_0(M_{Pl})$ & $\Lambda_1(M_{Pl})$ &${\Lambda_3}(M_{Pl})$ \\ [0.5ex] 
 \hline\hline
 1 &$ 0.021$ & $0.904 $&$ -0.15$ & $2.5\times 10^{-4}$&$ 8.4 \times 10^{-4} $&$8.19\times 10^{-4}$ &$2.32\times 10^{-4}$\\ 
 \hline
2 &$ 0.026$ & $0.774 $&$ -0.20$ & $2.5\times 10^{-4}$&$ 8.4 \times 10^{-4} $&$7.89\times 10^{-4}$ &$2.49\times 10^{-4}$\\ 
  \hline
\end{tabular}
\caption{Examples of parameters for the potential \eqref{eqpot}. }
\label{table1}
\end{table}
\end{center}

\begin{figure}[h!]
\centering
\includegraphics[width=85mm]{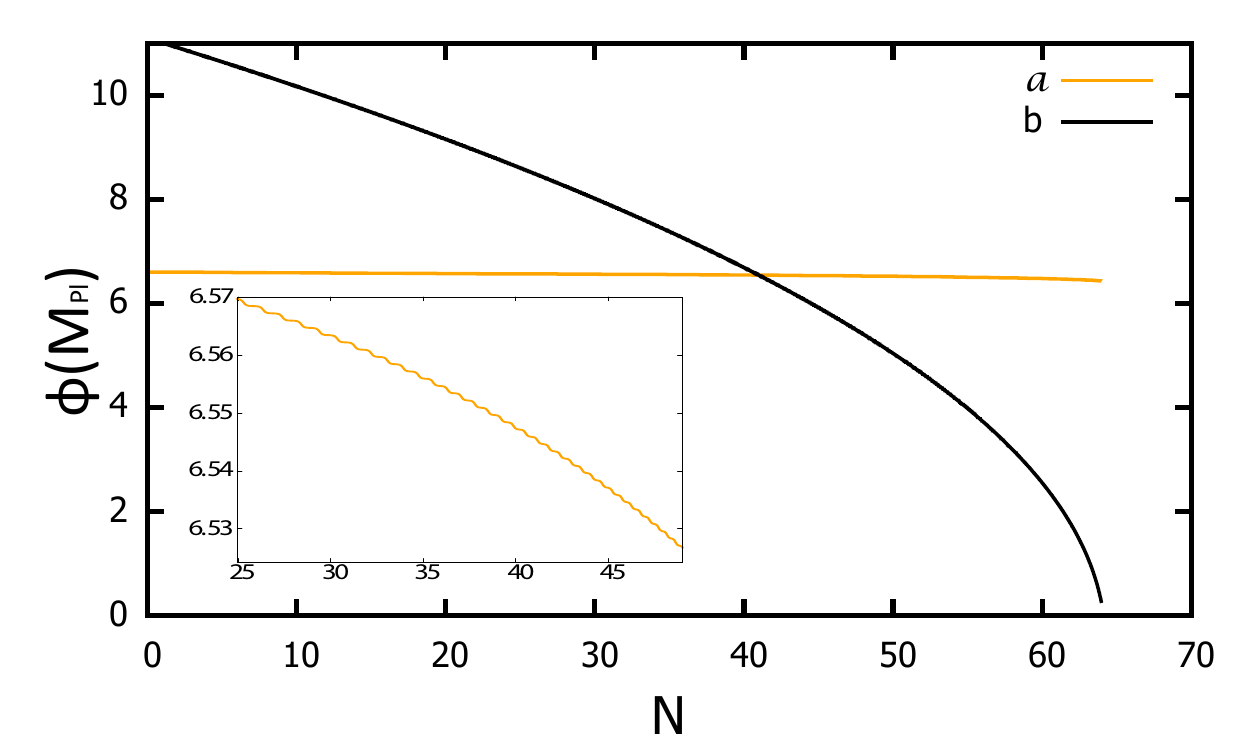}\includegraphics[width=85mm]{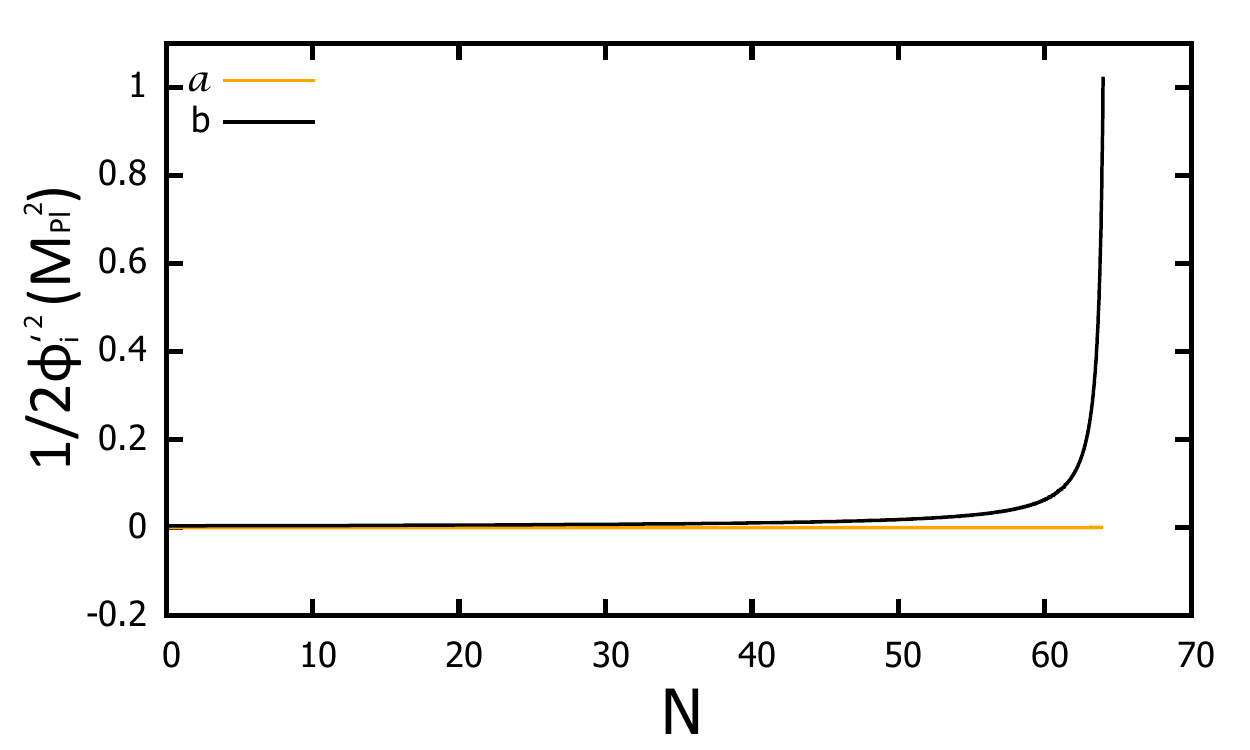}
\caption{Left: The evolution of the fields  $a,b$ for the first set of parameters in Table ~\ref{table1}. The inner plot shows the evolution of  field $a$. 
The oscillations, which appear as ``step"-like feature  in the evolution of the field $a$, are responsible for a similar oscillatory behaviour in the power spectrum.
Right: The quantity $\frac{1}{2} \phi_i'^2$ for field $a,b$.   The initial conditions are given in Table~\ref{table2}. }
\label{f00b}
\end{figure}

As one can see from Fig.~\ref{f00b} (left panel) the field $b$ is dominant during the first stage of inflation, when the oscillations of the field $a$ 
are suppressed compared to its linear term.  As time passes,  the value of $b$ decreases,  while the oscillations of the  field $a$ become  the leading effect. In the inner plot we show the evolution of the field $a$ (see Fig.\ref{f00}). The oscillations of the field appear effectively as 
many tiny ``step''-like patterns, however we stress here that  the evolution is smooth, and there are no discontinuities neither in the field nor in its first derivative.
 We expect this behaviour  to be reflected in the background of GWs~\cite{tetradis,Fumagalli:2021cel,Fumagalli:2021dtd}.

The numerical integration of Eq.~\eqref{21} stops when the   the slow-roll parameter $\epsilon_H$  reaches the value 1.  Although the $b $ field acquires smaller values than $a$,  as the number of efold increases (see Fig.~\ref{f00b} (left)), the square of its rate, $\dot b$, signals the end of the inflation. This claim is shown in Fig. \ref{f00b} (right), where we have plotted the contribution of each field to the parameter $\epsilon_H$ given in Eq.~ \eqref{22}.

At the region where the $a$ field is dominant, we expect these oscillations to be imprinted in the perturbations of the field. The evaluation of these perturbations is sketched in Appendix \ref{appendix1}. 
The power spectrum is 
\begin{equation}
P_R=\frac{k^3}{(2 \pi)^2}|{\mathcal{R}_k}|^2 \, ,
\end{equation}
where ${\mathcal{R}_k}$ is the curvature perturbation
\begin{equation}
\mathcal{R}_k=\Phi +  \sum_{\phi_i}  \frac{ \delta \phi_i }{\phi_i'}\, , 
\end{equation}
where ${\delta \phi_i }$ and $\Phi$ are the perturbation of the field and the Bardeen potential, respectively (see Appendix \ref{appendix1}). 
The detailed analysis of the evaluation of the perturbations, which we follow in the present work, is given in~\cite{Nanopoulos:2020nnh}.
The resulting power spectrum for the parameters in Table \ref{table1} is shown in Figs.~\ref{f01} and \ref{f02} (left panels). We observe that we have an important amplification of the spectrum at small scales, which is expected to be imprinted in the abundances of PBHs and in the energy density of the induced GWs.  The oscillations that appear in the spectrum were to be expected from to the ``step"-like features the field develops during its evolution~\cite{tetradis}, which are shown in the inner left  plot of Fig.~\ref{f00b}. 

\begin{figure}[h!]
\centering
\includegraphics[width=80mm]{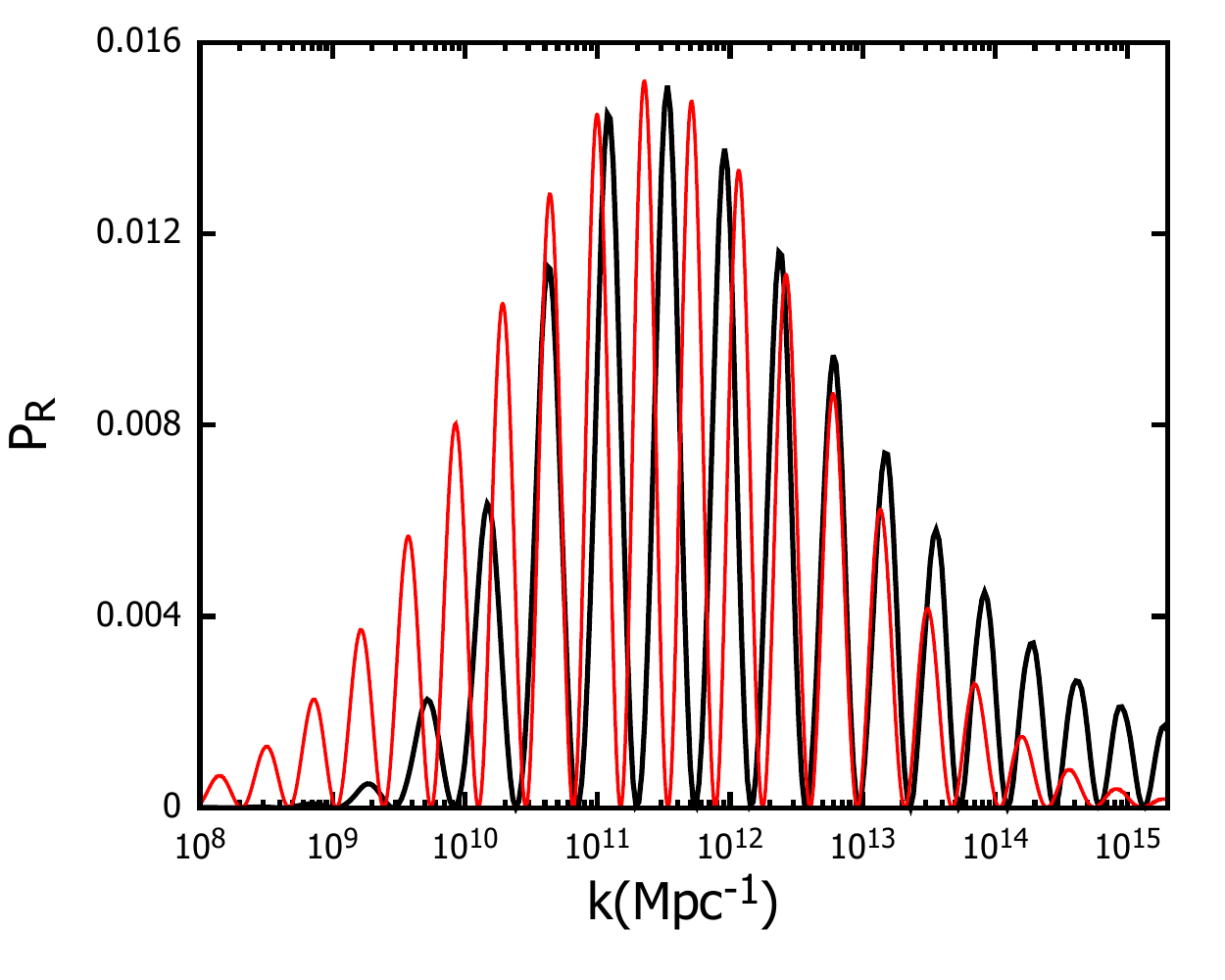}
\includegraphics[width=80mm]{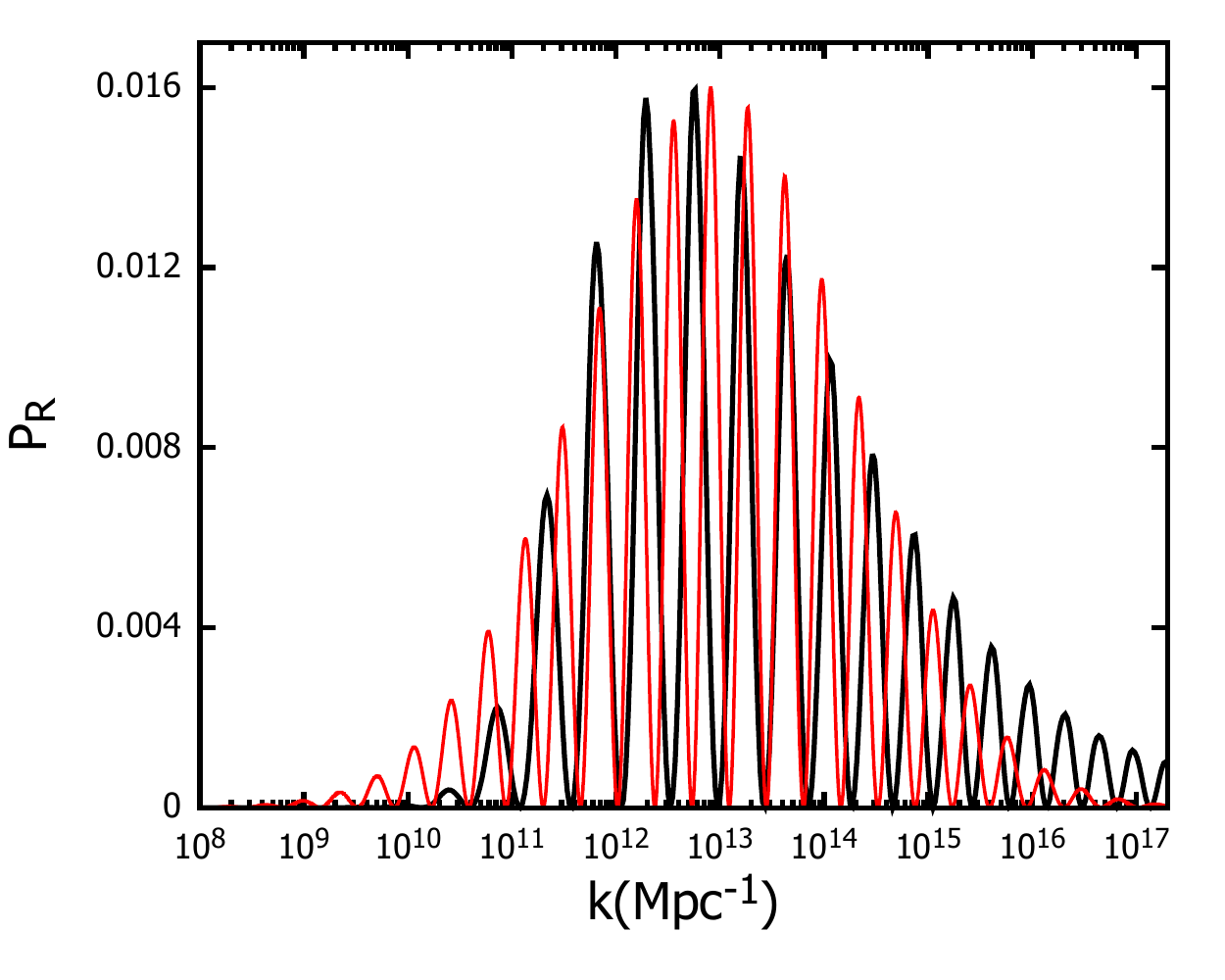}
\caption{ Left: The black line corresponds to the first set of Table \ref{table1}  and the red line corresponds to Eq.~\eqref{prfit} with the choice of parameters $\Delta=3$, $k_{\rm ref}=5\times 10^5$, $\vartheta_{l\rm og}=\pi/4$, $A_{\rm log}=1$, $\omega_c=4.77$, $\omega_{\rm log}=1.6\times \omega_c$ and $P_{max}=0.0076$.  Right:  The black line corresponds to the second set of Table \ref{table1}  and red line to Eq.~\eqref{prfit} with the choice of parameters $\Delta=3$, $k_{\rm ref}=3\times 10^6$, $\vartheta_{l\rm og}=\pi/4$, $A_{\rm log}=1$, $\omega_c=4.77$, $\omega_{\rm log}=1\times \omega_c$ and $P_{max}=0.0080$.}
\label{f07}
\end{figure}

We now remark that in ref.~\cite{Fumagalli:2021cel}, it was proposed that the oscillations in the  power spectra come in general into two categories, or a combination thereof, depending on whether the modulations of the power spectrum are  periodic functions of the scale $k$ or of the logarithm of $k$,  
\begin{align}\label{logk}
P_R (k) &= \overline P_R (k) \Big( 1 + A_{\rm lin} \, {\rm cos}(\omega_{\rm lin}\, k + \vartheta_{\rm lin})\Big)\, \nonumber \\
P_R (k) &= \overline P_R (k) \Big( 1 + A_{\rm log} \, {\rm cos}(\omega_{\rm log}\, {\rm ln}(k/k_{\rm ref}) + \vartheta_{\rm log})\Big)\,
\end{align}
where $k_{\rm ref}$ is a reference scale, $\overline P_R (k)$ is a smooth envelop, and the parameters $A_i$, $\omega_i, \vartheta_i$, $i=$lin, log
are phenomenological.  
In \cite{Fumagalli:2021cel} a specific form of the 
envelope function for the ln(k)-dependent oscillation spectrum has been proposed, which implies that the power spectrum assumes 
the following form:
\begin{equation}
P_{R} (k) = P_{max}\exp\left( -\frac{1}{2 \Delta^2} {\rm ln}\left(\frac{k}{k_{\rm ref}} \right)^2\right) \, \Big( 1 + A_{\rm log} \, {\rm cos}(\omega_{\rm log}\, {\rm ln}(k/k_{\rm ref}) + \vartheta_{\rm log})\Big)
\label{prfit}
\end{equation}
where $\Delta$, $P_{max}$ are parameters \cite{Fumagalli:2021cel}. In  \cite{Fumagalli:2021cel} it was demonstrated that the log(k)-dependent power spectra give rise to a resonant peak in the density of the GWs, provided the frequency $\omega_{\rm log}$ is sufficiently large, larger than a critical value 
$\omega_{\rm log} > \omega_c = 4.77$.   In Fig. \ref{f07} we plot the power spectrum from  Table \ref{table1} (black lines) and the $P_{R}/k_{\rm ref}$ from Eq.~\eqref{prfit} and we notice that they have a similar pattern. Thus, the enhanced densities of PBHs we discuss in the next subsection \ref{sec:pbhgw}
may be attributed to such resonant ln(k)-dependent modulations of the power spectrum with $O(1)$ coefficients $A_{\rm log}$. It goes without saying, however, that at present this discussion should only be viewed as indicative, given that detailed fits of our numerical spectra to (combinations of) the patterns \eqref{logk} are still pending, being reserved for a future publication. Moreover, and even more important, an analytic connection of ln($k$) dependences to the underlying  microscopic string models, whose world-sheet non-perturbative  effects lead to such patterns in the power spectra, is not yet available. Nonetheless, this does not affect the conclusions of the present work on the enhancement of PBHs and GWs, as a result of the periodic structures in the axion potentials \eqref{effpot} (or \eqref{eqpot}), which we discuss below.

Before proceeding to a discussion on the generation of PBHs and GWs,  we need to discuss the prediction of the observable constraints of inflation for the proposed potential \eqref{eqpot}.  The spectral index $n_s$ is calculated as 
\begin{equation}
n_s=1+\frac{d\ln P_R}{d\ln k}\, .
\label{ns}
\end{equation}
This calculation is performed near the pivot scale  $k_*=0.05Mpc^{-1}$.

Another useful observable quantity at CMB scale is the tensor-to-scalar ratio, which can be evaluated from:
\begin{equation}
r= \frac{P_T}{P_R},
\label{ratio}
\end{equation}
where ${P_T}$ is the power spectrum of tensor perturbations at scales where $k=\text{a(t)} H$ ($\text{a(t)}$ is the scale factor, not to be confused with the axion field $a$), which is given by:
\begin{equation}
P_T=\frac{2}{\pi^2}H^2.
\end{equation}

For the parameter sets given in Table \ref{table1}, one can find the predictions for $n_s$ and $r$ in Table~\ref{table2}.
\begin{center}
\begin{table}[ht]
\begin{tabular}{||c| c c c c|} 
 \hline
SET & $a_{ic}$ & $b_{ic}$ & $n_s$ &$r$ \\ [0.5ex] 
 \hline\hline
 1 &$ 6.605$ & $11.100 $&$0. 9638$ & $0.062$\\ 
 \hline
 2 &$ 4.932$ & $11.400 $&$0.9619$ & $0.060$\\ 
  \hline
\end{tabular}
\caption{Spectral index, $n_s$, tensor-to-scalar ratio, $r$, and the initial conditions for the $a$ and $b$ axion fields, for the corresponding sets of parameters given in Table ~\ref{table1}.}
\label{table2} 
\end{table}
\end{center}
The corresponding values for these quantities, which are evaluated at a pivot scale of $k_*=0.05Mpc^{-1}$, are  given from the  Planck collaboration~ \cite{Planck2018} 
\begin{equation}
  n_s =
    \begin{cases}
      0.9659 \pm 0.0041 & \text{Planck TT,TE,EE+lowEB+lensing}\\
      0.9651 \pm 0.0041 & \text{Planck TT,TE,EE+lowEB+lensing+BK15}\\
       0.9668 \pm 0.0037 &  \text{Planck TT,TE,EE+lowEB+lensing+BK15+BAO}
    \end{cases}       
    \label{ns-obs}
\end{equation}
and
\begin{equation}
r <
    \begin{cases}
      0.11 & \text{Planck TT,TE,EE+lowEB+lensing}\\
      0.061  & \text{Planck TT,TE,EE+lowEB+lensing+BK15}\\
       0.063  & \text{Planck TT,TE,EE+lowEB+lensing+BK15+BAO} \, .
    \end{cases}  
    \label{ratio-obs}
\end{equation}
Therefore, our  analysis is consistent with the basic  observable for inflation as measured by Planck collaboration. 
In the next subsection we calculate the  fractional abundance of PBHs and  the energy density of GWs.

\subsection{Primordial Black Hole formation and induced Gravitational Waves \label{sec:pbhgw}}

The enhancement of the power spectrum we have just discussed,  can be interpreted as leading to a significant amount of PBHs and GWs generated in the radiation epoch.
 We stress once more at this stage that, in the scenario of \cite{ms} adopted here, inflation is of  RVM-type~\cite{lima}, and is due to the non linearities of the cosmic energy density, specifically the terms proportional to the fourth power of the Hubble parameter, due to condensation of anomalies. The evolution of the $b$ axion field drives the slow-roll nature of this inflation.

In this subsection we show that this amplification can produce a large fraction of the DM in form of PBHs.
 Moreover we show that this amplification can be depicted in the future space-based experiment of GWs. The analysis we follow is similar to those of~\cite{Nanopoulos:2020nnh,inflectionpoint2}.  Before proceeding we would like to mention that the relation between between induced GWs and PBHs depends on the expansion history of the early universe, hence, it is in  general model dependent~\cite{domenech}. Indeed, in our context, we shall see below, that the results depend on the type of class of models  used, that is on the hierarchies of scales \eqref{constr}, \eqref{constr3}. 

\begin{figure}[h!]
\centering
\includegraphics[width=75mm,height=55mm]{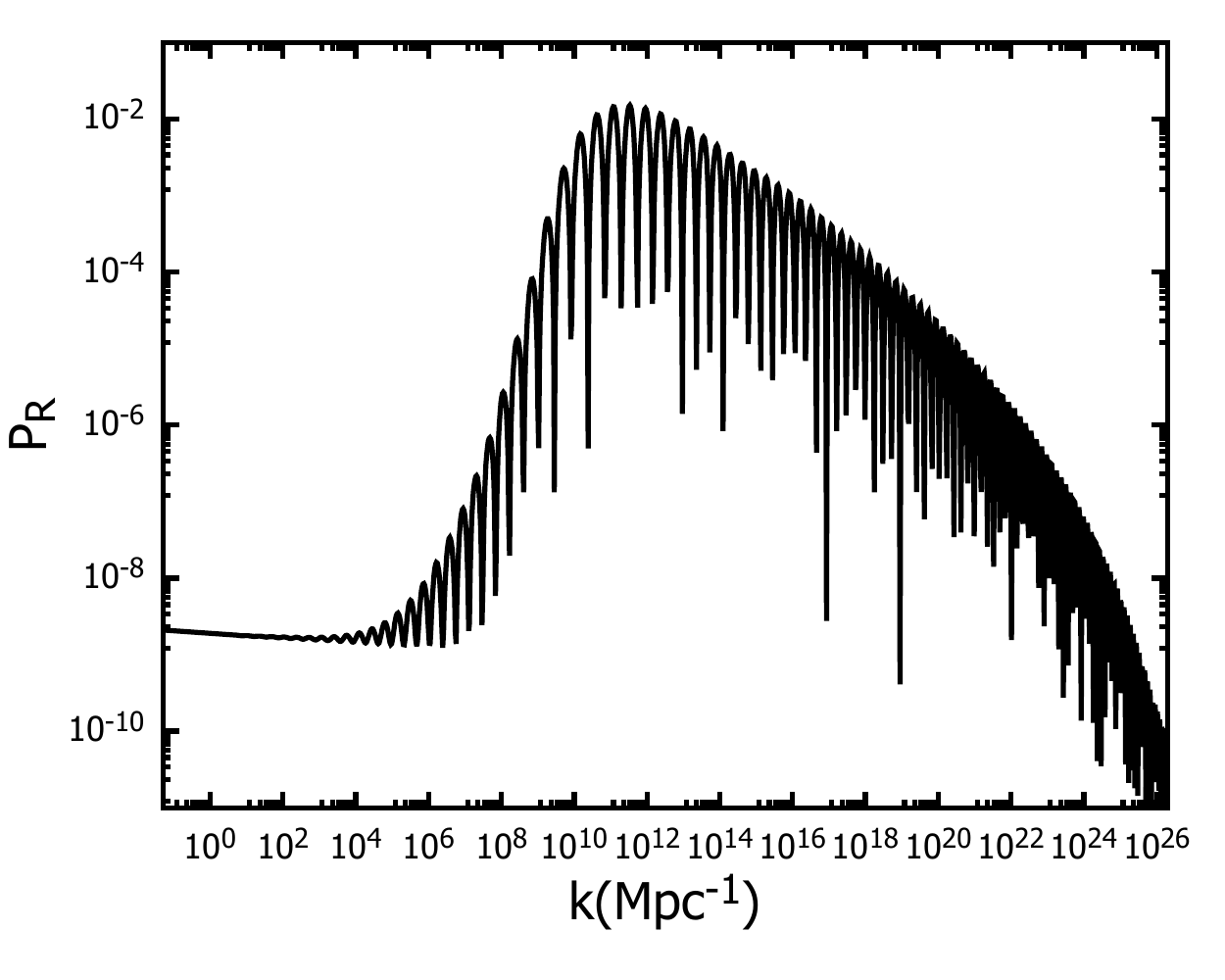}
\includegraphics[width=75mm,height= 63mm]{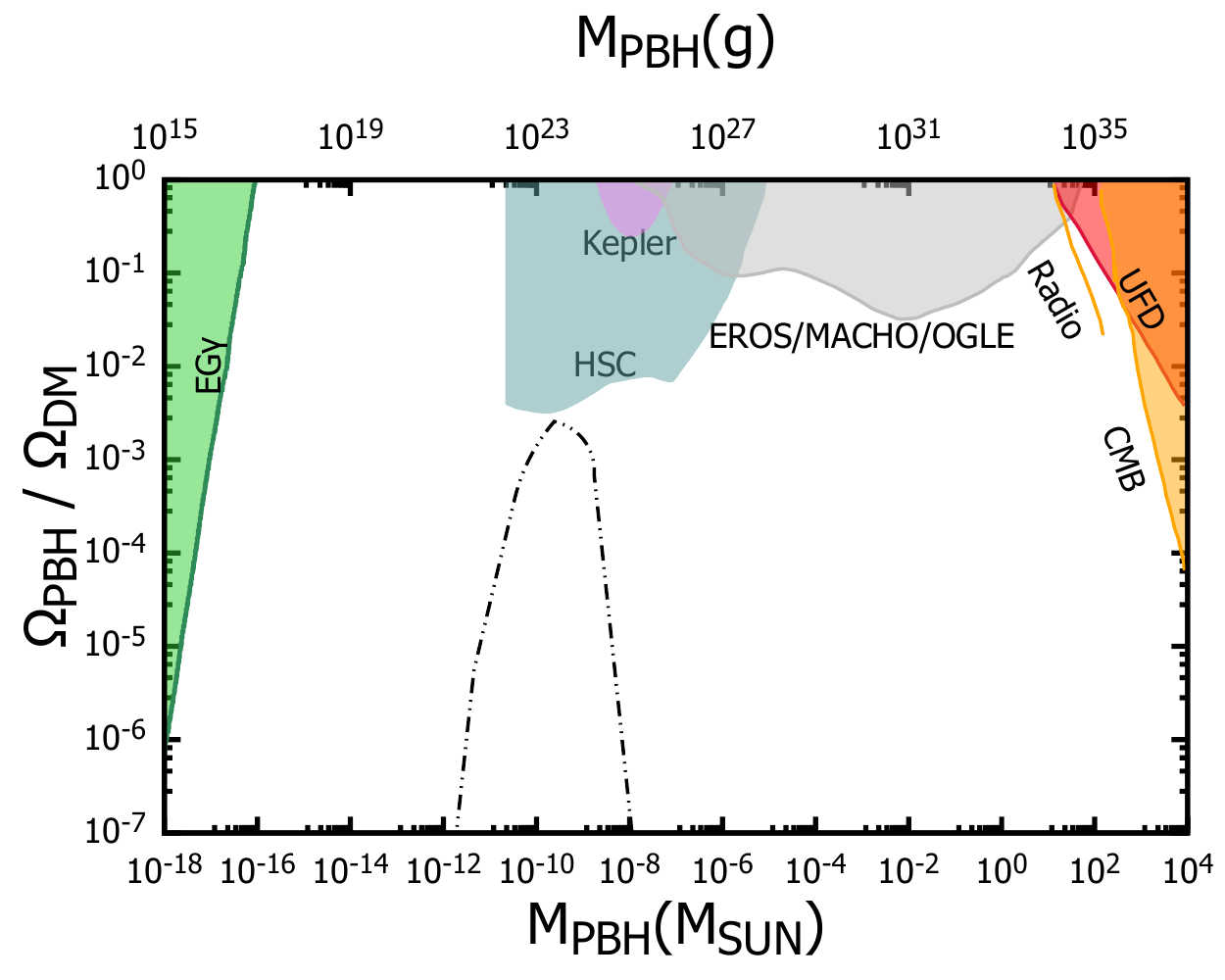} \\ \includegraphics[width=80mm]{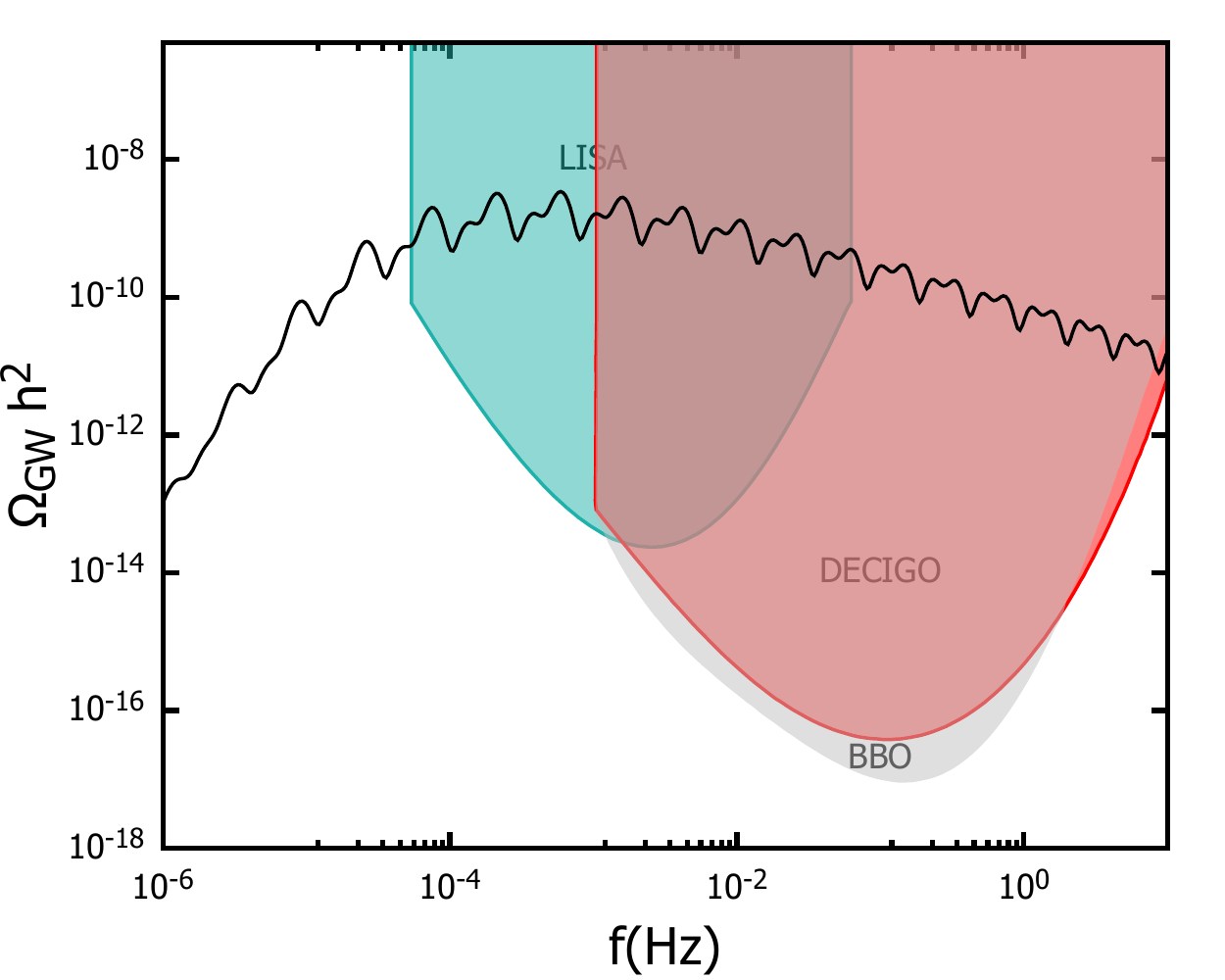}
\caption{ The results of power spectrum, fractional abundance of PBHs and the energy density of induced GWs for the first set of parameters given in Table \ref{table1}.  The initial conditions are  given in Table \ref{table2}. The fractional abundance of PBHs is  $f_ {PBH}=0.01$. }
\label{f01}
\end{figure}

\begin{figure}[h!]
\centering
\includegraphics[width=75mm,height= 55mm]{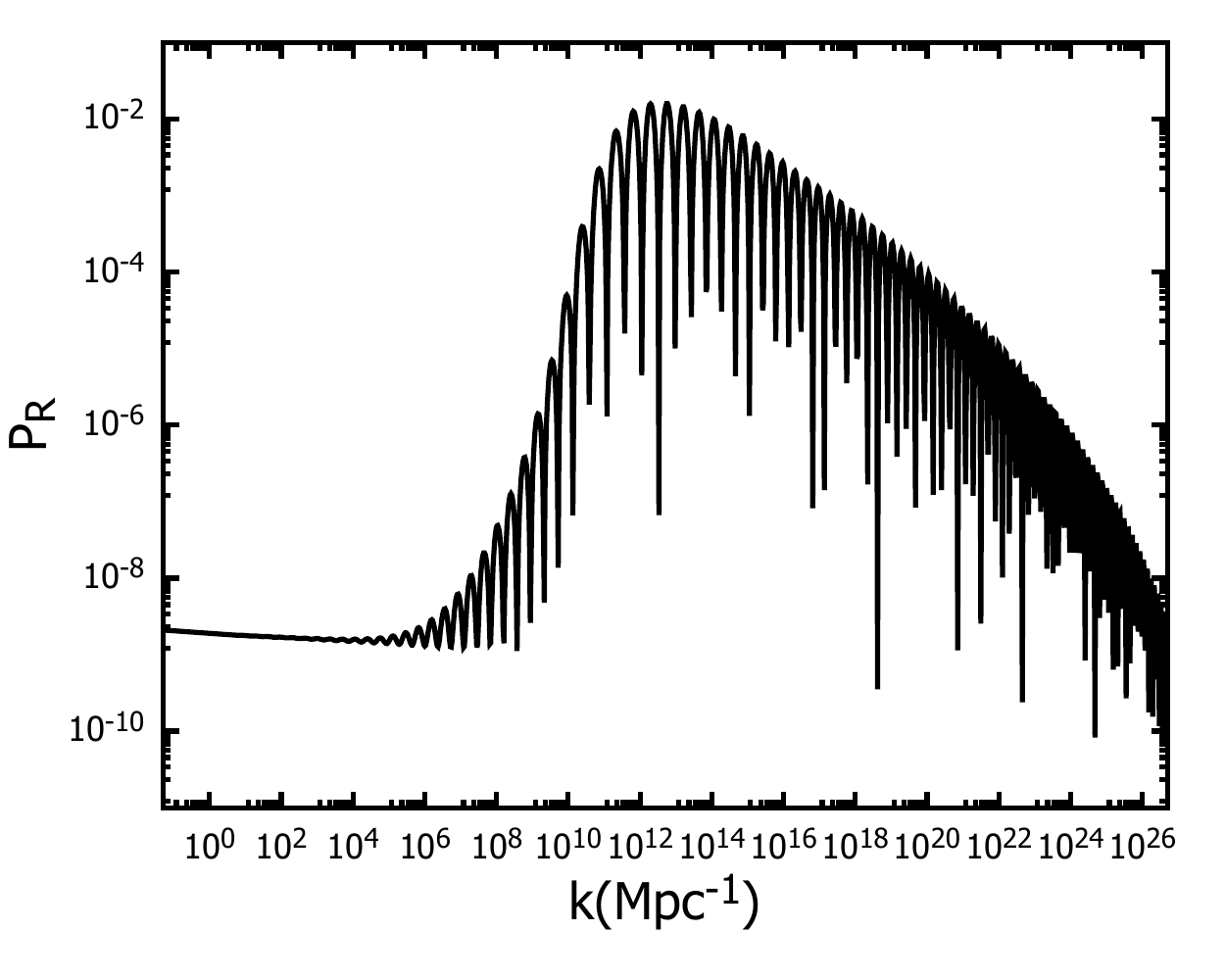}
\includegraphics[width=75mm,height= 63mm]{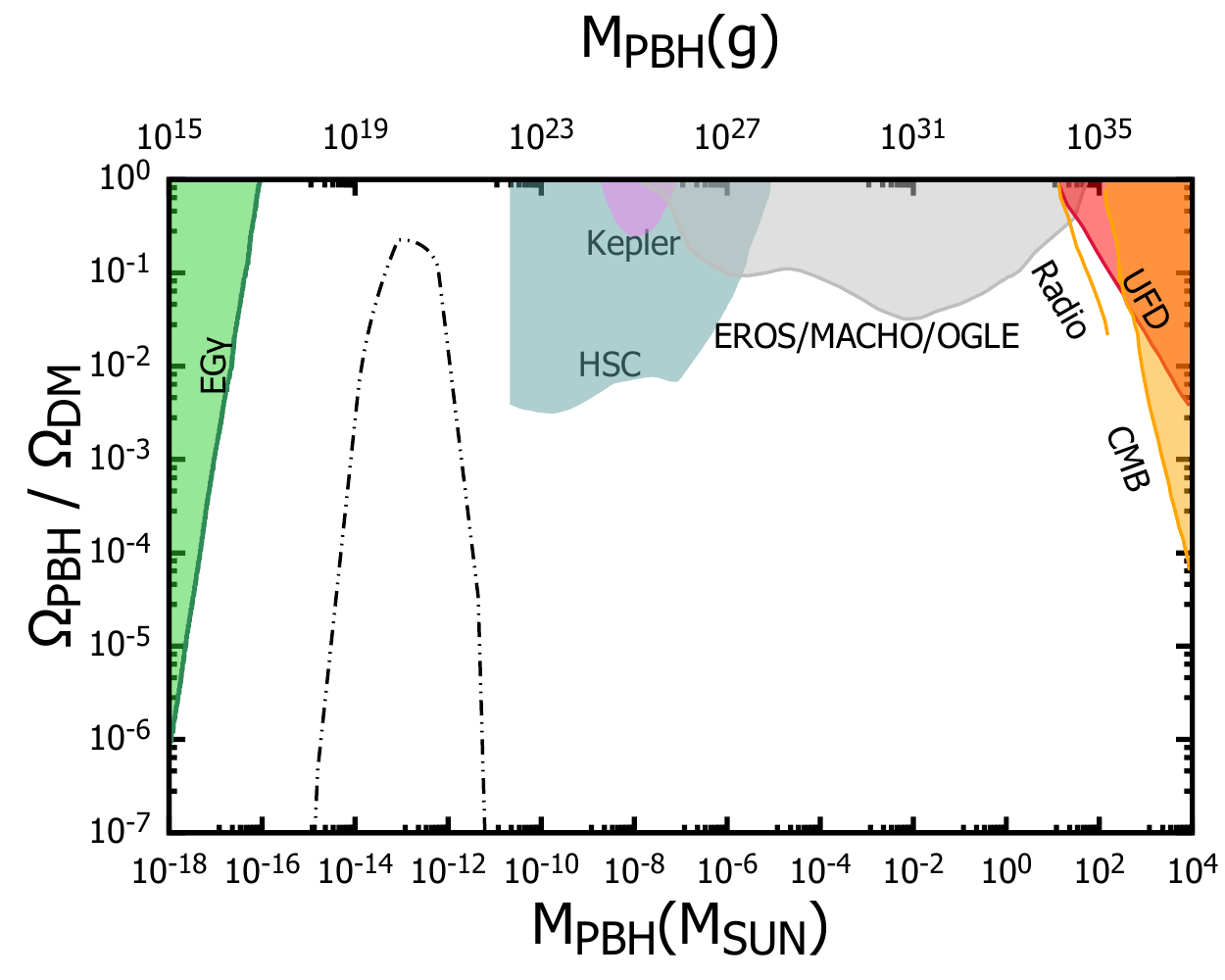}\\
\includegraphics[width=80mm]{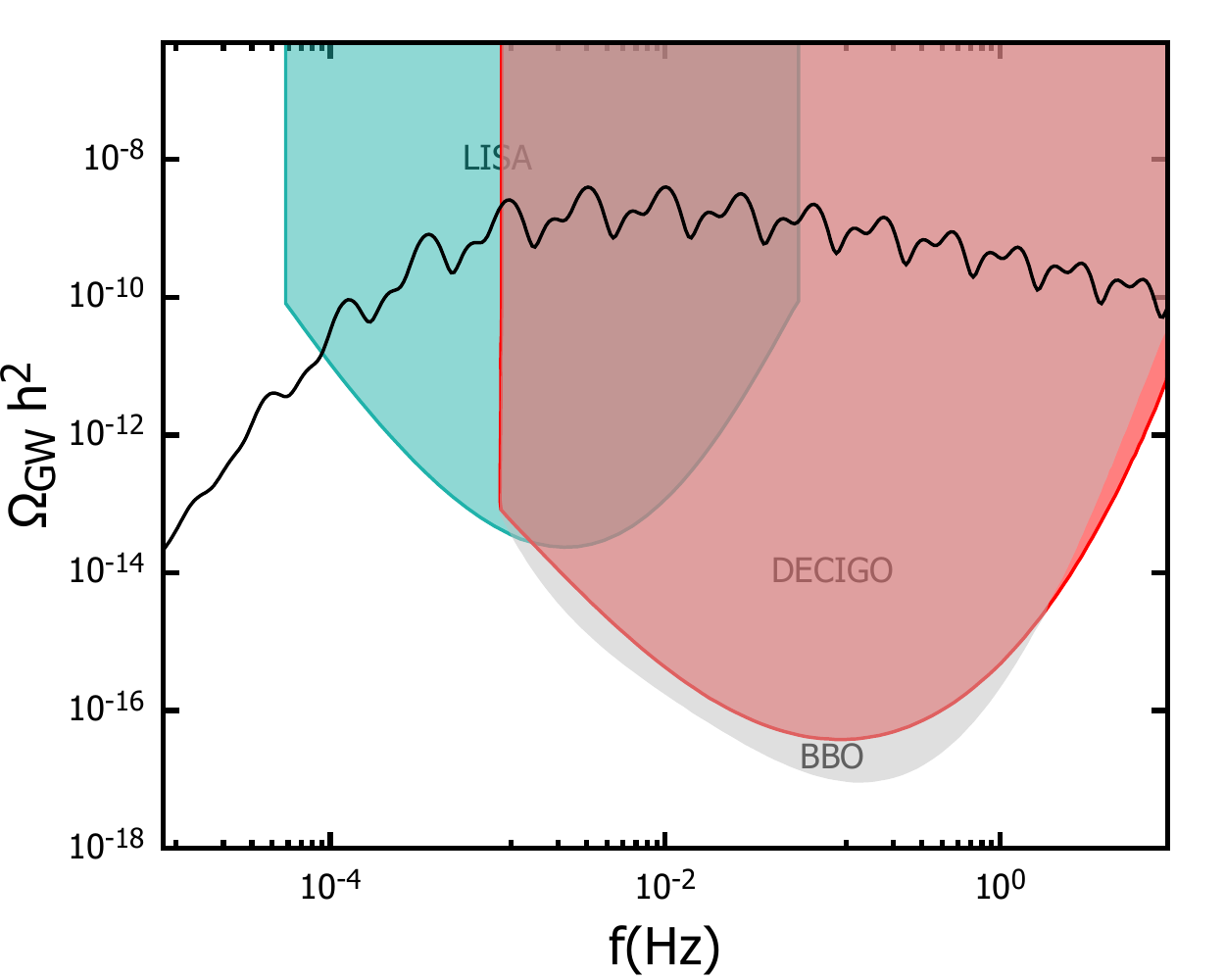}
\caption{The results of power spectrum, fractional abundance of PBHs and the energy density of GWs for the second set of Table \ref{table1}. The initial conditions are given in Table \ref{table2}. The fractional abundance of PBHs is  $f_{PBH}=0.80$.}
\label{f02}
\end{figure}

The present fraction of abundances of PBHs, $\Omega_{PBH}$,  with mass $M_{\mathrm{PBH}}$ over the total DM abundance, $\Omega_{DM}$ is \cite{Inomata:2017okj}
\begin{equation}
    \frac{\Omega_{PBH}}{\Omega_{DM}}(M_{\mathrm{PBH}}(k))=1.52 \times 10^8\left( {\frac{\gamma}{0.2}}\right)^{3/2}\left( \frac{g_*}{106.75} \right)^{-1/4} \left(\frac{M_{\mathrm{PBH}}(k)}{M_{\odot}}\right)^{-1/2} \beta(M_{\mathrm{PBH}}(k)) \, , 
 \label{opbh}
\end{equation}
where  $M_\odot$  is the solar mass (denoted as $M_{\rm SUN}$, in the figures), and $k$ is the comoving wavenumber. $g_*$ is the total number of relativistic degrees of freedom when PBHs are produced, which here we take to be its standard-model value, 106.75,\footnote{For the phenomenological purposes of our current work, we note that the model-dependent value of $g_*$, even taken in inimal supersymmetric standard model ($g_*=228.75$), 
can  affect  by at most 20\% the PBHs abundance.
 When the latter is embedded in string theory models, the $g_*$ can increase significantly (up to $10^3$ in extreme cases), but since $g_*$ is raised to the powers: $-\tfrac{1}{4}$ in Eq.~\eqref{opbh} and $-\tfrac{1}{6}$  in Eq.~\eqref{masspbh} (below), our results in this work will not be affected quantitatively by this.} $\gamma$ is a factor which depends on gravitational collapse and  we consider $\gamma \approx 0.2$ \cite{Carr:1974nx}. The parameter $\beta$ denotes the mass fraction of a Universe collapsing into PBHs of mass $M_{\mathrm{PBH}}$. It is obtained from the  integral of the overdensity $\delta$, which follows a Gaussian   probability
\begin{equation}
\label{42}
\beta(M_{\mathrm{PBH}}(k))= \frac{1}{\sqrt{2 \pi \sigma ^2 ( M_{\mathrm{PBH}}(k) )}} \int^{\infty}_{\delta_c} d\delta \,  \exp \left(  -\frac{\delta ^2}{2 \sigma^2( M_{\mathrm{PBH}}(k)) } \right) \, . 
\end{equation}

For the value of $\delta_c$ we take into consideration the studies~\cite{deltac,Musco:2020jjb}. 
The predicted abundance turns out to be  exponentially sensitive to this value. Generally in the 
literature different values for  $\delta_c$ are quoted~\cite{deltac}. 
Numerical simulations prove  that there is no unique
value for the threshold, because it depends on the shape of the power spectrum~\cite{deltac,Musco:2020jjb}.
In our analysis we have taken $\delta_c=0.4$.  
Moreover, $\sigma$ is the variance of the curvature perturbation,  which is  related to the comoving wavenumber $k$ as:
\begin{equation}
\label{40}
\sigma^2 \left(     M_{\mathrm{PBH}}(k)  \right)= \frac{16}{81}  \int \frac{dk' }{k'} \left(\frac{k'}{k}\right)^4 P_{R}(k') \tilde{W}\left(\frac{k'}{k}\right) ,
\end{equation}
and $\tilde{W}$ is a window function.  We assume that this function is the Gaussian distribution, following standard treatments. 
The mass of PBHs (expressed in grammars g) is given by \cite{inflectionpoint2}
\begin{equation}
 M_{\mathrm{PBH}}(k)=10^{18} \, \left( \frac{\gamma}{0.2}\right)\left( \frac{g_*}{106.75}\right)^{-1/6}\left( \frac{k}{7\times 10^{13} Mpc^{-1}} \right)^{-2}\, {\mathrm g}.
 \label{masspbh}
\end{equation}

Hence, the fractional abundances of PBHs, evaluated from \eqref{opbh}  for the two sets of parameters of Table \ref{table1}, are given in the middle panels of Figs.~\ref{f01},\ref{f02}. In these figures  we also display
the excluded regions from various  studies \cite{Carr:2009jm}.
The present abundance of PBHs can be evaluated as~\cite{inflectionpoint2} 
\begin{equation}
f_ {PBH} = \int dM_{\mathrm{PBH}}(k) \, \frac{1}{M_\mathrm{PBH} (k)} \, \frac{\Omega_ {PBH} }{\Omega_{DM}}( M_{\mathrm{PBH}}(k))\, ,
\label{fofDM}
\end{equation}
In this equation we substitute $M_{\mathrm{PBH}}$ from  \eqref{masspbh} and the ratio ${\Omega_ {PBH} }/{\Omega_{DM}}(M_{\mathrm{PBH}})$ from  \eqref{opbh}. Then, we evaluate the integral in  equation \eqref{fofDM}  over the comoving wavenumber $k$ in the range  shown in the left panels of Figs.~\ref{f01}, \ref{f02} (plots of power spectrum $P_R$).
The values of $f_ {PBH}$ for the sets in Table \ref{table1} are given in the captions of Figs.~\ref{f01}, \ref{f02} and are also summarised, for the reader's convenience, in Table \ref{table3}. 
 We remark at this point that the maximum value of $f_ {PBH}$ depends on the peak of the power spectrum and should respect  the allowed regions depicted in the middle plots of Figs.~\ref{f01}, \ref{f02}. We stress that  lighter PBHs with mass less than $10^{-17}M_{\odot}$ are expected to have entirely evaporated by now due to Hawking radiation~\cite{Hawking:1974rv}.
 As one observes, a significant fraction of DM   can be explained as being due to PBHs, as a consequence of the enhancement mechanism presented here. Such considerations compliment those of the string-inspired RVM models of \cite{ms}, in which another potential component of DM could come from the stringy axions~\cite{bms}, thus contributing to the rich cosmology of the latter~\cite{axiverse,marsh}.

Next,  we calculate the amount of the induced GWs produced during the radiation dominated era.
 The amount of GWs is evaluated from second-order (tensor)  perturbations. Nevertheless, the second order perturbations are related to the first order perturbations \cite{tensortoscalar}. 
Therefore, the energy density of the GWs is evaluated from the scalar power spectrum  \cite{Ananda:2006af,Espinosa:2018eve}:
\begin{equation}
\Omega_{GW}(k)=\frac{\Omega_r}{36} \int^{\frac{1}{\sqrt{3}}}_{0} d \mathrm{d} 
\int ^{\infty}_{\frac{1}{\sqrt{3}}} \, d s \left[  \frac{(s^2-1/3)(\mathrm d^2-1/3)}{s^2-\mathrm d^2}\right]^2 \, 
P_{R}(kx)P_{R}(ky)(I_c^2+I_s^2),
\label{o1}
\end{equation}
where ${\Omega_r}$ is the density of radiation today and its values is ${\Omega_r = 5.4  \times 10^{-5}}$.
The variables $x$ and $y$ are defined as
\begin{equation}
x= \frac{\sqrt{3}}{2}(s+ \mathrm d), \quad  y=\frac{\sqrt{3}}{2}(s-\mathrm d).
\label{eq4.2}
\end{equation}
Finally, the functions $I_c$ and $I_s$ are given by the equations
\begin{gather}
I_c=-36 \, \pi \, \frac{(s^2+\mathrm d^2-2)^2}{(s^2-\mathrm d^2)^3} \, \Theta(s-1)\\
I_s=-36 \, \frac{(s^2+\mathrm d^2-2)^2}{(s^2-\mathrm d^2)^2} \Bigg[ \frac{(s^2+\mathrm d^2-2)}{(s^2- \mathrm d^2)} \, \ln \left| \frac{\mathrm d^2-1}{s^2-1} \right| +2 \Bigg] \, .
\label{eq4.3}
\end{gather}

In Figs.~\ref{f01}, \ref{f02} (right panels), we present the GWs energy density, as a function of the frequency, for the corresponding sets given in 
 Table~\ref{table1}. 
 We notice  that the oscillation pattern in the power spectrum is imprinted in the abundances of GWs as well \cite{tetradis}. 
Moreover, we remark  that this proposed model can explain the future observations of LISA, DECIGO and BBO \cite{Audley:2017drz,Sato:2017dkf,Yagi:2011wg}. In these figures we also present the expected results of these future experiments for reasons of comparison. The reader should also recall the indicative fit of the power spectra to ln($k$) dependences ({\it cf.} Fig.~\ref{f07}), which may lead to resonant peaks in the GWs spectra~\cite{Fumagalli:2021cel}, that can lead to in principle detectable effects in future interferometers~\cite{Fumagalli:2021dtd}.

\section{A model with a different hierarchy of scales \label{sec:sas}} 

In this section we present, for completeness, an alternative string-inspired scenario, which is also characterised by a significant enhancement in the scalar power spectrum, but corresponds to the hierarchy of scales \eqref{constr3}.  Such a class of models have been studied in Ref.~\cite{sasaki}, and we discuss it here for comparison with our RVM model described in the previous sections, which is based on the hierarchy \eqref{constr} (or, equivalently, \eqref{constrl3}). 
For sufficiently large $\Lambda^4_2/(f_a \, \Lambda^3_0) > 1$,
the parameter $g_1 $ could also be taken to be larger than unity
\begin{align}\label{constr2b}
g_1 =  \frac{f_b}{f_a} + \frac{\Lambda^4_2}{f_a\, \Lambda^3_0} \,\, {>} \,\,1\, , 
\end{align}
a feature that characterises the models studied in \cite{sasaki}. In this section we show that under a proper choice of parameters of the potential \eqref{effpot}, we can obtain a significant enhancement of the scalar power spectrum and the density of the PBHs produced during inflation. 

\begin{figure}[h!]
\centering
\includegraphics[width=75mm,height= 55mm]{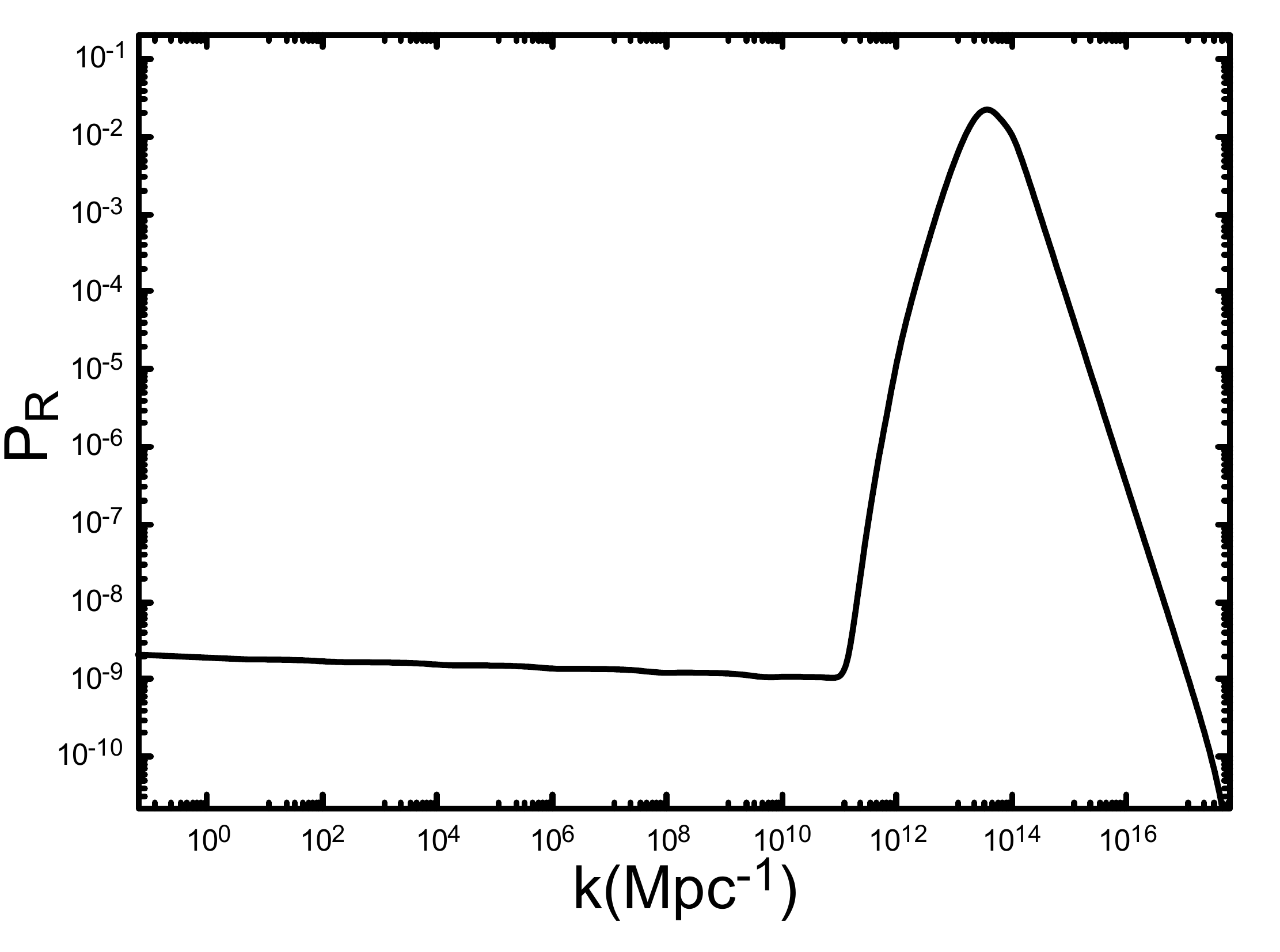}
\includegraphics[width=75mm,height= 63mm]{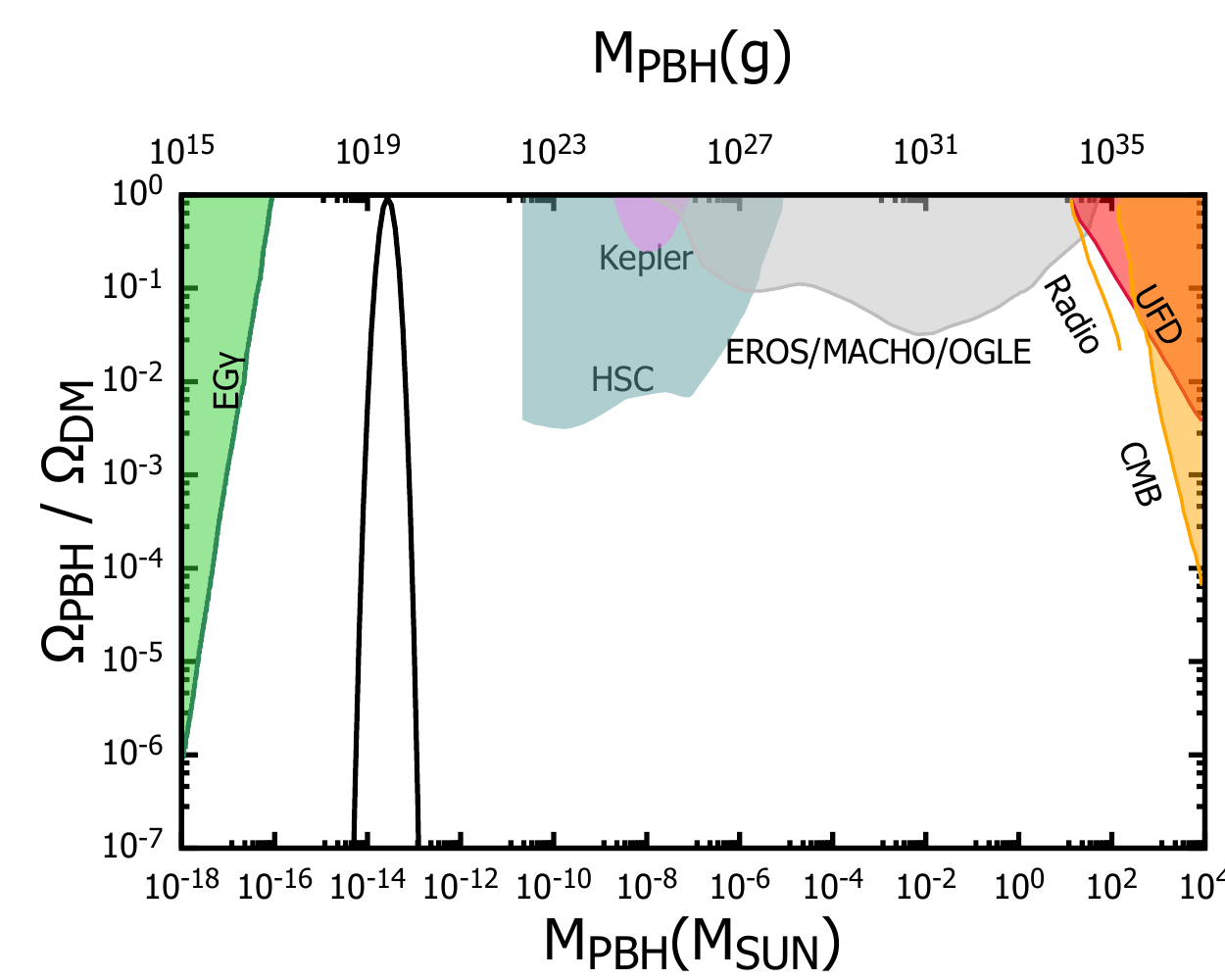}\\
\includegraphics[width=80mm]{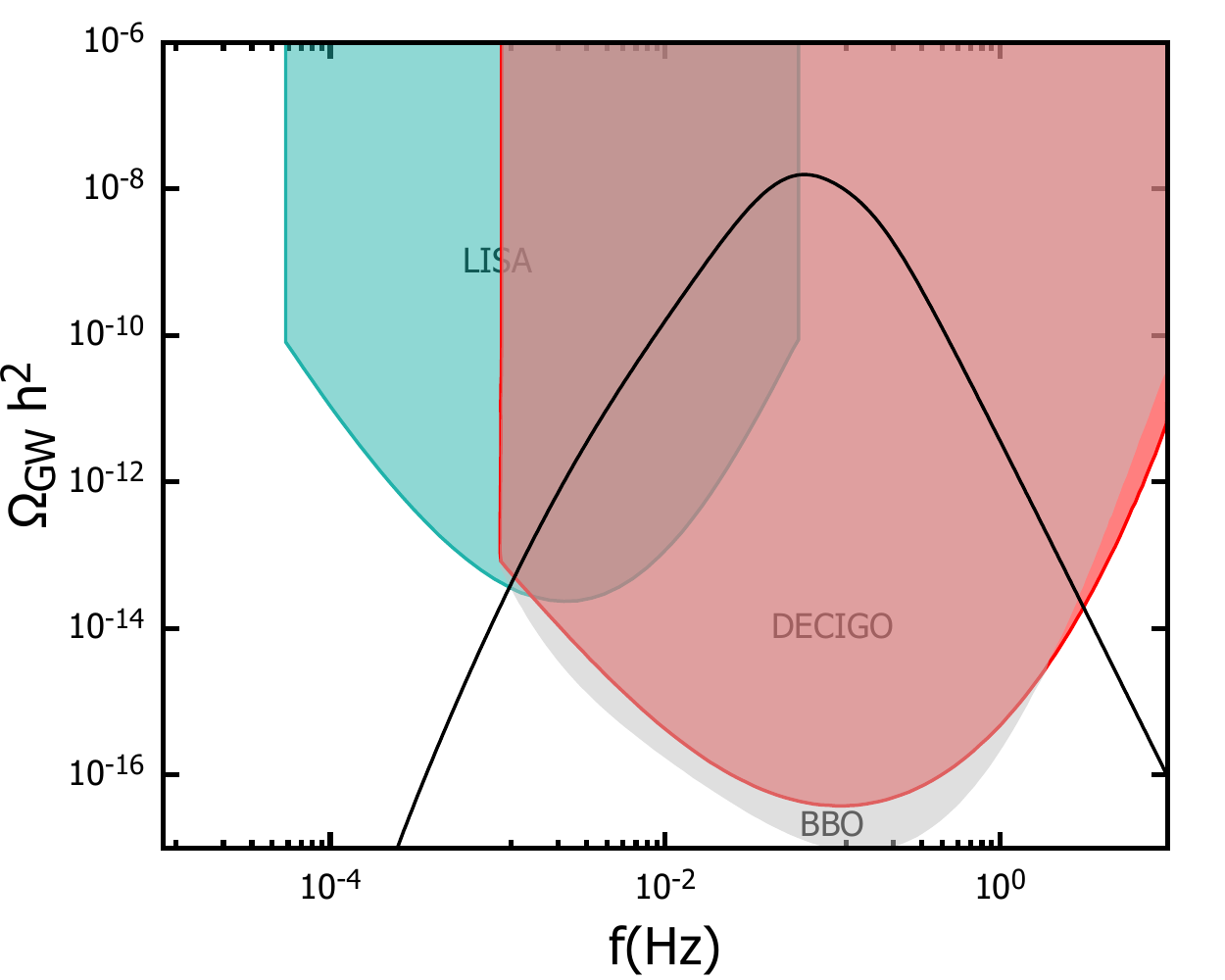}
\caption{Upper left panel: The  power spectrum.  Upper right panel: The abundances of PBHs. Lower panel: The energy density of GWs 	All diagrams are for the choice of parameters (\ref{choice}).  The initial conditions are $(a_{ic},b_{ic})= (7.5622, \, 0.5220)$.}
\label{f06}
\end{figure}

In this class of models, there is a first inflationary stage, driven by the brane-compactification axion field $a$, which follows the axion-monodromy patterns of \cite{silver}, which is then followed by an inflationary period driven by the KR field $b$, as a result of its linear potential term due to the anomaly condensate~\cite{ms}. In this scenario, the field $b$ prolongs inflation, which can be viewed as a second stage inflationary era. Par contrast, as we discussed previously, in the model of \cite{ms}, it is {\it not} the axion field that leads to inflation, but the {\it non-linearities}
 of the RVM vacuum energy.   The linear axion potential serves as a provider of the slow-roll nature of the axion fields in that case.\footnote{We note though that the second inflationary era could still be of RVM type, should one wish to embed such a class of cosmology models into the stingy RVM framework of \cite{ms}.}
Nevertheless, from the point of view of the potential amplification of GWs perturbations and enhancement of the densities of PBHs, the microscopic origin of inflation is not relevant.

 The choice of parameters for the potential \eqref{eqpot} which 
 respects the scale hierarchy \eqref{constr3} is as follows:
\begin{equation}
\Lambda_0=8.4\times 10^{-4} M_{\rm Pl},  \quad  \Lambda_1=9.7 \times 10^{-3} M_{\rm Pl},     \quad g_1=110,\quad g_2=1.779\times
10^4,\quad \xi=-0.09,\quad f=0.09 \,\, M_{\rm Pl}\,.
\label{choice}
\end{equation}
With this choice, we indeed have: $\Lambda_1> \Lambda_0$.  
    
 In this case we consider a different mechanism for the amplification of the scalar power spectrum at small scales, than the one we followed in previous sections whereas relation \eqref{constr} was valid. 
Specifically, we follow the method of a near inflection point in the scalar potential~\cite{inflectionpoint,inflectionpoint2}.  Indeed, under a proper choice of parameters, the form of the axion potential \eqref{eqpot} in this case allows  the existence of an inflection point. We stress that these models have the advantage of naturally having such a feature without further modifications~\cite{Ballesteros:2019hus}.  We remark that in~\cite{sasaki}  the enhancement of a similar potential, which  follows relation \eqref{constr3}, occurs via the relaxation mechanism analyzed in \cite{Graham:2015cka}. In particular in  \cite{sasaki} the corresponding  field $a$ has oscillations that become dominant  as time passes. Hence,  those oscillations lead to an amplification of the scalar power spectrum.  However, in the present work, we choose an appropriate set of parameters which implies a near inflection point  of the effective scalar potential.
 We note that  the first and second derivatives of the potential in the direction of the field $a$ can be approximately zero for a proper choice of parameters defining a near inflection point \cite{inflectionpoint,inflectionpoint2}. It is in this sense,  that the enhancement mechanism we propose here is different from that of \cite{sasaki}. 
       
In order to proceed to the calculation of the power spectrum, we evaluate the equations of motion ~\eqref{21} and the equations for the perturbation of the fields given in Appendix \ref{appendix1}, as before. The resulting power spectrum  is given in the left panel of the Fig.~\ref{f06}. Repeating the same methodology as in the previous case, we evaluate the fractional abundance of PBHs and the energy density of GWs. The results are given in the  middle and right panel of Fig.~\ref{f06}, respectively. We notice that in this case we can obtain an important amplification, which can explain the production of PBHs and GWs~\cite{sasaki}. 
The fraction of DM due to PBHs, $f_ {PBH}$ ({\it cf.} \eqref{fofDM}), is given in Table \ref{table3}, where we also present the value of the peak of the power spectrum and the mass of PBHs.   In that Table, for comparison, we also give the values of these quantities for the two sets of parameters presented in Table \ref{table1}, for the models of \cite{ms}, obeying the scale hierarchy \eqref{constr}.

 At this point one can notice the apparent difference in the morphology of the enhanced power spectra between Figs.~\ref{f01},\, \ref{f02} and Fig.~\ref{f06} 
(left panels). For the case of the GACS model with the hierarchy of scales \eqref{constr}, the resulting power spectrum 
 exhibits an intensive oscillating pattern, due to the presence of the $\cos(a/f_a)$ in the potential. In this term, the axion coupling $f_a$ gets a small value resulting  to the 
 amplification of the oscillations. Moreover, the whole $\cos(a/f_a)$ term becomes significant at later times, due to scale hierarchy. 
 This oscillating pattern leaves its imprint in the energy density of the GWs. 
This pattern is not reflected on the PBHs abundance, because the peak there is predominantly
monochromatic Fig.~\ref{f06} , while in Figs.~\ref{f01},\,\ref{f02} the spectrum is sufficiently wide~\cite{tetradis}. 

Par contrast, in the class of models characterised  by the scale hierarchy \eqref{constr3}, the
enhanced power spectrum is smooth. This is due to the fact that the amplification of the power spectrum results  from an inflection point in the 
scalar potential. Thus,  this smooth enhancement is reflected to the GWs energy density spectrum.  
 Consequently, future observations in the GWs spectra,  like those form LISA~\cite{Audley:2017drz},  can in principle distinguish between  these two classes of inflationary 
 models, depending, of course, on the sensitivity of the corresponding observations \cite{Fumagalli:2021dtd}.

We stress once again, that the maximum value of $f_ {PBH}$ depends on the peak of the power spectrum and the specific value of the $\delta_c$, as
it was previously discussed. 
Moreover,  it  should respect the allowed regions depicted in the middle plots of Figs.\ref{f01}-\ref{f06}, in agreement with the lower mass bounds $M_{\rm PBH} \gtrsim 10^{-17}M_{\odot}$,  due to the Hawking evaporation process~\cite{Hawking:1974rv}.
The restricted regions shown in these figures come from measurements of the extragalactic gamma-ray background, microlensing for Subaru (HSC), Eros/Macho/Ogle and Kepler,  dynamical heating of ultra faint dwarf (UFD), CMB measurements and radio observations \cite{Carr:2009jm}.

\begin{center}
\begin{table}[h]
\begin{tabular}{||c| c c c |} 
 \hline
SET & $P_{R} ^{peak}$ & $M_{PBH} ^{peak}(M_{\odot})$&$f_{PBH}$ \\ [0.5ex] 
 \hline\hline
 1 &$1.466 \times 10^{-2}$ &$2.394\times 10^{-10}$ &$0.009 $\\ 
 \hline
 2 &$1. 365 \times 10^{-2}$ & $8.313\times 10^{-14}$&$0.799 $\\ 
  \hline
 3 &$2.24  \times 10^{-2}$ &$1.791\times 10^{-14}$&$0.762 $\\ 
  \hline
  \end{tabular}
\caption{The value of the peak of the power spectrum, the mass of the corresponding PBHs  (in solar mass $M_{\odot}$ units), and the present-era abundance of PBHs for the two sets of parameters presented in Table \ref{table1} (sets 1, 2), respecting the scale hierarchy \eqref{constr} (string-inspired RVM cosmology \cite{ms}),  
and for the  choice of parameters given in   Eq.~\eqref{choice} (set 3), which obeys the scale hierarchy \eqref{constr3} (string-axion-monodromy class of models \cite{sasaki}).}
\label{table3}
\end{table}
\end{center}

The prediction for the observables $n_s$ and $r$ is evaluated from Eqs.~ \eqref{ns} and \eqref{ratio} and their values are given as follows:
 \begin{equation}
n_s=0.9623 ,\quad r =0.038.
\end{equation}
Therefore, this  model is  consistent~\cite{sasaki} with the observable constraints from Planck~\cite{Planck} .
 As in the case of the RVM models of \cite{ms}, this class of cosmological models  predicts the presence of PBHs. 
 From this point of view, therefore, these two classes of models behave similarly. 
 However, as we have already mentioned, their microscopic origin of inflation is entirely different. 

\section{Reheating Issues in the two classes of Models \label{sec:reh}}

 In our previous discussion we have assumed instantaneous reheating, which however may not be a universal feature of our models. 
Before concluding, therefore, we would like to discuss in this section the reheating mechanisms that could characterise the two classes of models discussed above, the stringy-RVM and the conventional string/brane axion monodromy model
and see under which circumstances a prolonged reheating period can be discarded. 
As we have already mentioned, the PBH densities and thus the spectrum of GW does depend on the thermal history of the post-inflationary Universe, and thus, understanding reheating in our models is an important aspect that needs clarification, and, moreover, helps the reader in comparing the physical differences between the two classes of the above-described string-inspired axion cosmologies. 

Generically, there are two ways by means of which a slow reheating process can affect the PBH mass function: 
\begin{enumerate} 

\item The first concerns the case of a slow reheating process, in which an intermediate mass-dominated era with equation of state $w=0$ precedes the radiation era, which may lead to a significant enhancement of the PBH mass fraction~\cite{carr} that plays the r\^ole of a DM in the post-inflationary Universe. 

\item The second way is due to the fact~\cite{kumar}  that the presence of a non-instantaneous reheating stage changes the mapping of different length scales from horizon exit to re-entry. This will affect  the normalisation of the scalar power spectra at the pivot scale and therefore the mass fraction, even for the monochromatic case (this last feature was missed in the analysis of \cite{Cai}, which was corrected by the work of \cite{kumar}). 
In this more general case, the equation of state of the fluid during reheating does not have to be that of matter. Although the latter leads to the most significant enhancement, other types of the equation of state also lead to significant enhancement of the fraction of PBHs that can play the r\^ole of DM, as the analysis of \cite{kumar} demonstrated for a few characteristic (albeit phenomenological) cases.

\end{enumerate}

Let us now examine whether any of the above can characterise the two classes of models we discussed previously.

In the second class of stringy axion monodromy models, studied in section \ref{sec:sas}, reheating is driven after the second inflation, where the axion $b$ dominates. For such models the reheating is instantaneous, and can be achieved by a standard quadratic (mass) terms for the axion~\cite{sasaki}. This was assumed in our analysis above. 

In the first class of models, the stringy RVM~\cite{ms}, inflation is driven by the (dominant) fourth-power of the Hubble parameter, that characterises the vacuum energy density $\rho^{\rm RVM}_{\rm vac}$,  due to the formation of the anomaly condensate \eqref{cond}: 
\begin{align}\label{rvmener}
\rho^{\rm RVM}_{\rm vac}  \simeq 3\, \kappa^{-4} \, \Big[ -1.65 \times 10^{-3} \, (\kappa H)^2 
+ \frac{\sqrt{2}}{3}|\overline b(0)|\, \kappa \,  \times 5.86 \times 10^6 \, (\kappa\, H)^4 \Big] + \dots \, ,
\end{align}
where $\overline b(0)$ is the value of the KR axion field at the onset of the RVM inflation, and the negative coefficient of the (subdominant) $H^2$ term is due to the contributions of the gravitational anomaly to the stress energy ternsor~\cite{ms}. The $\dots$ in \eqref{rvmener} denote the (subleading as far as inflation is concerned) contributions from the quantum fluctuations of the KR $b$ axion, and the compactification axion $a$ and their non-perturbative periodic potential, as appearing in \eqref{effpot}, which play no important r\^ole in our subsequent discussion on reheating. The reader should notice that 
\eqref{rvmener} pertains to the leading terms contributing to the ground state 
of the string vacuum, and this includes the background configuration of the KR axion field, which results in the appearance of the $\overline b(0)$ term in \eqref{rvmener}. The full linear KR axion $b$ potential \eqref{effpot} is recovered by replacing $\overline b(0)$ by the field $b(t)$.

\begin{figure}[h!]
\centering
\includegraphics[width=0.6\textwidth]{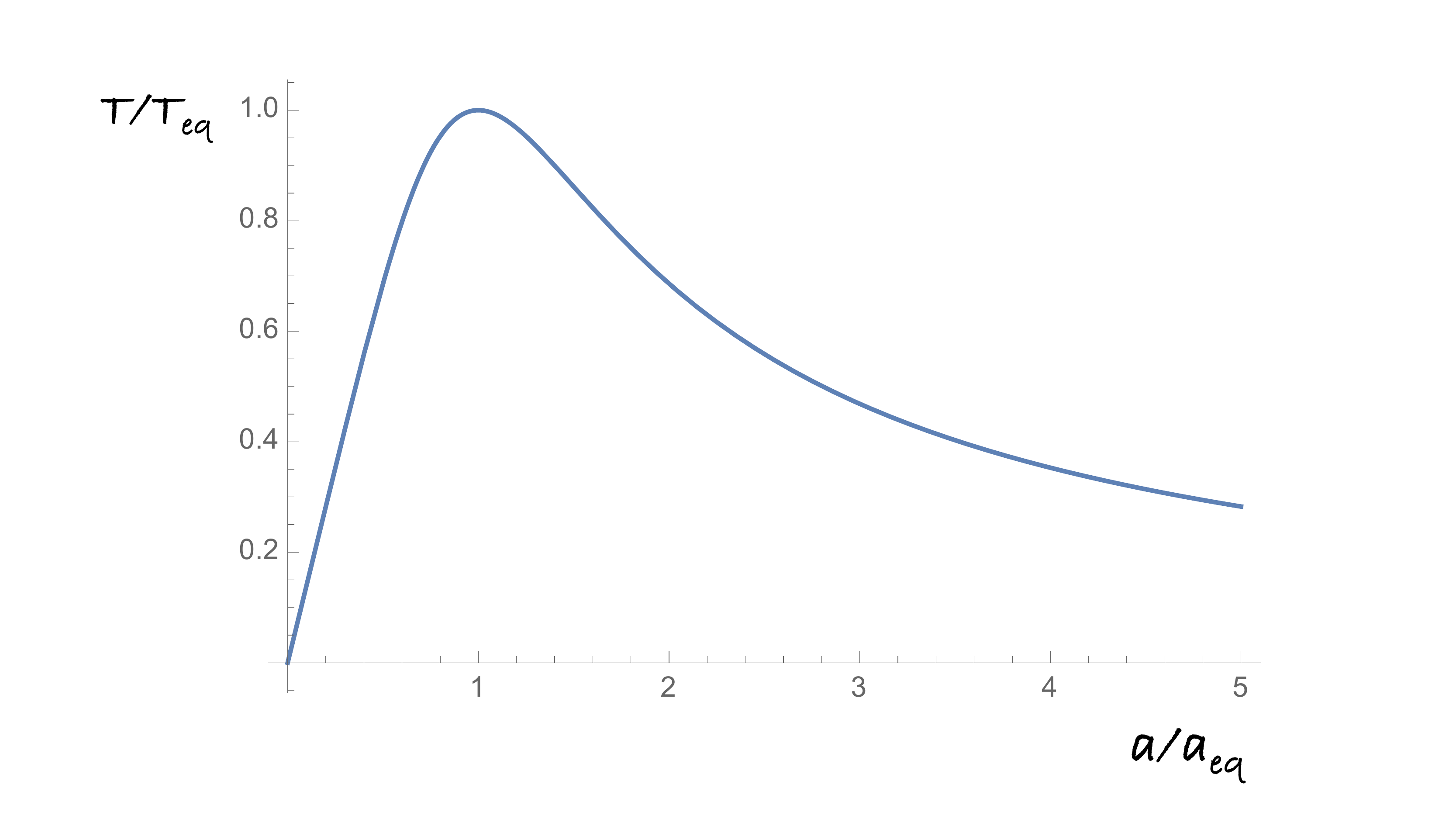}
\caption{ The function T (in units of $T_{\rm eq}$) versus the scale factor $\text{a}$ (in units of $\text{a}_{\rm eq}$) of the RVM universe, \eqref{radmod}. The suffix ``eq'' denotes the point of equality of the energy densities of vacuum and radiation (not to be confused with the 
standard-cosmology point of matter-radiation equality). The maximum value of $T$ is attained at that point, at which the scale factor $\text{a}=\text{a}_{\rm eq}$~\cite{lima}.}
\label{fig:Tvsa}
\end{figure}

The form \eqref{rvmener} is of a RVM type, for which we know that it has important implications for the thermal history of the Universe. Indeed, from the analysis in \cite{thermal}, we observe that there is a {\it prolonged adiabatic period} of reheating of the RVM Universe, rather than an instantaneous process, as happens in standard cosmology~\cite{kolb}.
The radiation particles appear as a result of the decay of the running de Sitter vacuum, which is metastable.  As discussed in \cite{thermal}, although the energy density during radiation obeys the usual $T_{\rm rad}^4$ scaling with the cosmic temperature of radiation era $T_{\rm rad}$, the standard 
cooling law of the Universe during radiation $\text{a}_{\rm rad}\sim 1/T_{\rm rad}$ (where $\text{a}_{\rm rad}$ is the scale factor, $\text{a}$, during the radiation dominance) is modified:
\begin{align}\label{radmod}
T_{\rm rad} =  \Big( \frac{90\, H_I^2 \, 
\kappa^{-2}}{\pi^2 \, g_{\star, \rm rad}} \,\Big)^{1/4} \, \frac{\Big(\text{a}/\text{a}_{\rm eq}\Big)}{\Big[1 + \Big(\text{a}/\text{a}_{\rm eq}\Big)^4\, \Big]^{1/2}} =  \, \sqrt{2} \, T_{\rm eq} \, \frac{\Big(\text{a}/\text{a}_{\rm eq}\Big)}{\Big[1 + \Big(\text{a}/\text{a}_{\rm eq}\Big)^4\, \Big]^{1/2}} \, , 
\end{align}
where the suffix ``eq'' indicates the point at which the running vacuum energy density equals that of radiation, and we have defined $T_{\rm eq} = T_{\rm rad}(\text{a}_{\rm eq}) = \Big(\frac{45\, H_I^2}{2 \, \pi^2\, \kappa^2 \, g_\star}\Big)^{1/4}$, with $H_I$ the RVM inflation scale, and $g_\star$ the degrees of freedom of the created massless modes of the cosmological model at hand (see Appendix \ref{sec:appRVM} for some brief review of the basic concepts and techniques which lead to \eqref{radmod}). The function T versus a
\eqref{radmod} is plotted in fig.~\ref{fig:Tvsa}.

 From \eqref{radmod} we observe that $T_{\rm eq}$ is the {\rm maximum} attained reheating temperature. 
The result \eqref{radmod} is a consequence of the cosmic evolution of the RVM framework~\cite{lima,thermal}, which our stringy RVM model~\cite{ms} also obeys.
One peculiarity of the latter model is that the end of inflation coincides with the cancellation of the primordial gravitational anomalies, that dominate and drive the RVM inflation in the model by the gravitational anomalies induced by the chiral fermions that are generated at the end of the inflationary period as a result of the decay of the stringy RVM vacuum. This process is not instantaneous but occurs during the aforementioned prolonged reheating period.

The important feature of \eqref{radmod} is the fact that for $\text{a} \ll \text{a}_{\rm eq}$ one observes a phase of the expanding Universe in which the temperature grows linearly with the cosmic scale factor, attaining a maximal value at vacuum-radiation equality $\text{a}=\text{a}_{\rm eq}$ (see fig.~\ref{fig:Tvsa}). 
Thus, instead of having the usual instantaneous (highly-non adiabatic) reheating~\cite{kolb}, the RVM Universe exhibits a prolonged non-equilibrium heating period, as a result of the decay of the vacuum, which drives progressively the Universe into the radiation phase.
This linear-$T$  phase is responsible for the entropy production in the RVM Universe~\cite{thermal}. 
For $\text{a} \gg \text{a}_{\rm eq}$ the Universe reaches a perfect fluid adiabatic phase, during which the temperature decreases in the usual way following a $T^{-1}$ cooling law for the Universe scale factor. 

In the standard RVM scenarios~\cite{thermal}, despite the long reheating process, inflation is succeeded by the radiation era, with no matter-domination intervention. 
This is what we assumed in our analysis in the previous sections. 
 However,  in our stringy RVM~\cite{ms}, discussed in this work, one needs  a more careful look, before a definite conclusion is reached concerning the absence of an intermediate  matter-dominance phase before radiation dominance. Indeed,  
the presence of string compactification axions in our model, with periodic sinusoidal modulations induced by  non-pertrurbative
(world-sheet instanton) effects, may, under some circumstances, change the above conclusion. In fact, as observed from \eqref{effpot}, these instanton effects lead to quadratic (mass) terms in the vacuum energy density:
\begin{align}\label{massterms}
V(a, b) \ni \frac{1}{2}\, \frac{\Lambda_1^4}{f_a^2}\, a(t)^2 
\end{align}
implying an axion mass of order 
\begin{align}\label{axionmass}
m_a \sim \Lambda_1^2/f_a\,. 
\end{align} 
Depending on the values of the parameters (which, in turn, depend on the details of the underlying microscopic string model), one might encounter a situation, in which, as a result of the presence of the terms 
\eqref{massterms}, at the end of the RVM inflation, the axion field $a(t)$ oscillates slowly around its minimum value, and reheats the Universe, in parallel with the decay of the RVM vacuum itself.  In case of an early dominance of the axion-induced reheating process, which, we stress, depends on the details of the parameters of the model, one might have a rather long epoch of an early matter dominance before the radiation era, which slowly reheats the Universe, as for instance 
was assumed in \cite{carr}. Such a matter-dominated reheating will be succeeded by the (also slow) radiation-era reheating that characterises the decay of the RVM vacuum (once photons are created by the decay of the RVM vacuum at the  end of the RVM inflation era, as assumed in the scenario of \cite{ms}), then the axions couple to the  chiral gauge anomaly 
$\propto \frac{1}{2\, f_a} \, \varepsilon^{\mu\nu\alpha\beta}\,  a(t)\, F_{\mu\nu}\, F_{\alpha\beta}$, with $F_{\mu\nu}$ the Maxwell tensor, in case of Abelian anomalies, or the non-Abelian field strength tensor in case of chiral anomalies of non-Abelian gauge groups. The massive axions will then decay to massless radiation modes for photon pairs, for instance, $\gamma$ the corresponding (tree-level) decay width will be of order 

\begin{align}\label{widthaxion}
\Gamma (a \rightarrow \gamma\, \gamma) \sim \frac{m_a^3}{64\pi \, f_a^2} \sim \frac{\Lambda_1^6}{64\, \pi \, f_a^5}\,,
\end{align}
where we used \eqref{axionmass}. 
For our stringy RVM class of models, in both sets of parameters of Table \ref{table1}, 
we have in order of magnitude 
$\Lambda_1 \sim f_a $, and thus 
\begin{align}\label{width}
\Gamma \sim \frac{\Lambda_1}{64\pi} \sim 10^{-6}\, M_{\rm Pl}\,, 
\end{align}
implying very short life times for the decay of the axion into radiation,  of order $10^6 \, t_{\rm Pl}$, where $t_{\rm Pl} = M_{\rm Pl}^{-1}$ (in units $\hbar =c =1$) is the reduced Planck time. The width $\Gamma $ \eqref{width} is compatible with the upper bound 
$\Gamma \le 10^{14}\, {\rm GeV}$ required for 
instant reheating of the Universe~\cite{Planck}, as assumed in our previous analysis. Moreover, given that 
during inflation $H \sim 10^{-5}\, M_{\rm Pl}$, the condition for an intermediate matter dominated era before  radiation $\Gamma \ll H$ is not satisfied for the set of parameters used in section \ref{sec:pbhgw} ({\it cf.} Table \ref{table1}).  Indeed, such a condition characterises models in which $\Lambda_1 \ll \Lambda_0$, and not simply $\Lambda_1 < \Lambda_0$ as we considered above.
Therefore, the computation of the production of PBHs in the radiation era, which we considered in our analysis in the previous sections, is correct for the set of parameters we have used. 

There are ambiguities however in string models, due to the fact that the axion does not decay only to standard model massless particles, but also to exotic massless modes, e.g. from hidden sector of the underlying theory. This is a problem that we do not address in the current phenomenological study. Nonetheless, it has been addressed in the string/brane literature~\cite{blumen}, in which models with suppressed decays of the axion to hidden sector particles as compared to those of the standard model have been explicitly constructed. Thus, we may state that, in the context of our first class of models,  associated with the stringy RVM framework, one does not have an intermediate mass-dominated reheating phase, before the decay of the vacuum to radiation-dominated eras.

In such a case, therefore, the second way  of enhanced PBH mass fraction, due to the prolonged reheating process of the stringy RVM vacuum,  which affects in general the involved scales, especially the pivotal one used in the estimates, 
might be in operation, as per the discussion in \cite{kumar}, \cite{Cai}.
This would make our estimates in section \ref{sec:pbhgw} on the PBH fraction that act as DM  rather conservative. Let us discuss this issue from the point of view of the stringy RVM framework, in which, as we shall discuss below, one may avoid altogether long reheating periods, by appropriate choosing the relevant set of parameters.

Indeed, in  such this case, the modified relation \eqref{radmod}, between radiation temperature and scale factor of the Universe, which is responsible in general for a prolonged reheating, will contribute in general to the modification of the relevant scales involved, and thus to the overall values of the PBH fraction that could play the r\^ole of a DM component in post-inflationary epochs. In our estimates of the PBHs densities above, we have used sets of parameters for which the reheating process is of very short duration. However, in general, if one has a prolonged RVM reheating process,  one may have an increased fraction of PBH that plays the r\^ole of DM, given that the mass function of the PBH is affected by such a prolonged RVM reheating process. 
In this case, one has (in the expressions below, for clarity, we reinstate the scale factor today $\text{a}_0$ explicitly):
\begin{align}\label{reheat}
\frac{\text{a}_i}{\text{a}_0} = \frac{\text{a}_i}{\text{a}_{\rm end}} \, \frac{\text{a}_{\rm end}}{\text{a}_{\rm reh}}\, 
\frac{\text{a}_{\rm reh}}{\text{a}_0}\, ,  \qquad 
 \frac{\text{a}_i}{\text{a}_{\rm e}} = e^{-N_I}\, \quad \frac{\text{a}_{\rm end}}{\text{a}_{\rm reh}} = e^{-N_{\rm reh}}\,, 
\end{align}
where $\text{a}_0$ denotes the scale factor today, $\text{a}_i$ denotes an initial scale factor, at the beginning of the RVM inflation, $\text{a}_{\rm end}$ is the scale factor at the end of the RVM inflation, $\text{a}_{\rm reh}$ is the scale factor during the reheating process, $N_I > 0$ is the number of e-foldings during the RVM inflation, and $N_{\rm reh} > 0$ denotes the number of e-foldings during reheating. For the RVM scenarios, including the stringy RVM~\cite{ms} of interest to us here, the equation of state of the fluid during reheating is that of radiation $w_{\rm reh} = 1/3$, but $N_{\rm reh} \ne 0$ in general, depending on the parameters of the model.

As discussed in Appendix \ref{sec:appRVM}, the evolution equation of the dominant radiation fluid is \eqref{evol}, with the rate of change of the term on the right-hand side 
being estimated as: 
\begin{align}\label{evolRVMdetail}
\dot \rho_{\rm RVM} = 3\, \kappa^{-2} \, \Big( H^2 \, [ 2 \, \nu \,\frac{\dot H}{H} + \dot \nu ] + 
\frac{H^4}{H_I^2} \, [ 4 \, \alpha \, \frac{\dot H}{H}  + \dot \alpha ] \Big)
\end{align}
In the standard RVM, leading to \eqref{radmod}, the coefficients $\nu$ and $\alpha$ are assumed constant. In the stringy RVM~\cite{ms}, though, as mentioned above, the generation at the end of inflation of extra chiral matter which has its own anomaly terms implies that the latter can eventually cancel the primordial gravitational anomalies. This is captured effectively by introducing a decreasing with time $\alpha$ coefficient in front of the $H^4$ term in \eqref{rvmener}. Moreover, as shown in \cite{ms}, the generation of cosmic electromagnetic (and in general radiation-relativistic) fields at the end of inflation leads to terms $H^2$ in the energy density with a {\it positive} coefficient, which therefore imply initially a decrease in magnitude of the negative coefficient of the $H^2$ terms in \eqref{rvmener} due to the primordial gravitational anomalies. All these features are captures by introducing an appropriate cosmic time dependence of the RVM coefficients $\nu$ and $\alpha$ in the RVM expression for the vacuum energy density in \eqref{evol}, which leads to \eqref{evolRVMdetail}. In our assumption in previous sections was that this time dependence was such that the rate of change of the RVM energy density 
 at the end of inflation reduced to  zero much quicker than in the standard, non stringy, 
 RVM case, thus justifying our {\it instantaneous reheating approximation} that lead to the PBHs abundance estimates in section \ref{sec:2mon}.  
  From the point of view of the stringy RVM framework, one may say that we choose the parameter $\mathcal D$, which appears in the Hubble evolution \eqref{evolHsol} (see Appendix \ref{sec:appRVM}), and depends on microscopic parameters in string theory, that affect the boundary conditions of the beginning of the stringy RVM inflation~\cite{ms}, 
 in such a way that $a_{\rm eq} = \frac{1}{\mathcal D}$ ({\it cf.} \eqref{Daeq}) 
is such that  the linear-in-$T$ reheating period in fig.~\ref{fig:Tvsa} is much smaller than the radiation period. This would imply instantaneous reheating.

However, we mention for completion that,depending on details of the underlying string theory microscopic model, 
there could be situations in the stringy RVM~\cite{ms}, including our axion models studied here, in which the modified coefficients of the $H^2$ and $H^4$ terms in \eqref{evolRVMdetail} vanish during a prolonged reheating period, before the system 
goes into the radiation phase. In such a case, during that prolonged reheating 
we might have, therefore:
\begin{align}\label{prolongcoeff}
[ 2 \, \nu \,\frac{\dot H}{H} + \dot \nu ] \simeq 0, \qquad 4 \, \alpha \, \frac{\dot H}{H}  + \dot \alpha \simeq 0\,.
\end{align} 
implying that the right-hand-side of the evolution equation \eqref{evol}, \eqref{evolRVMdetail}, would vanish during this prolonged reheating period, and the system would correspond to a perfect fluid with an 
equation of state $w_{\rm reh}$, and an energy density $\rho_{\rm reh}$ obeying the standard 
scaling \eqref{scalreh}. 
In that case, the analysis of \cite{kumar}, as described briefly in Appendix \ref{sec:appRVM}, can be applied in order to study the effects of such prolonged reheating on the PBHs densities. 
The dominant reheating period could 
be identified with a cosmic fluid with equation of state that of radiation, 
with the understanding though that it describes the prolonged reheating phase, during which there is still inflation characterised by $N_{\rm reh}$ e-folding at the end of the reheating phase.

 In the conventional RVM framework~\cite{lima} where the vacuum decays into radiation, the equation of state $w_m$ in \eqref{evol} of the dominant fluid during the prolonged reheating will be that of radiation, $w_{\rm rad} = w_{\rm reh} =1/3$,
but $N_{\rm ref} \ne 0$ in the non-instantaneous reheating case. 
It should be stressed that the modified radiation-temperature equation \eqref{radmod} cannot be used as such to describe this situation.  
In order to determine the
initial condition $\text{a}_i$ that we can use for our estimates of the primordial spectrum and PBH densities one needs to consider the procedure outlined in \cite{kumar} (and \cite{Cai}), which is reviewed in Appendix \ref{sec:appRVM}.

 Indeed, if one assumes that the entropy of the Universe is almost constant from the onset of radiation till the present era, which is also valid in the (stringy) RVM, where most of the entropy is produced during the adiabatic reheating, then standard manipulations~\cite{kumar} yield:
\begin{align}\label{scaling}
\frac{\text{a}_i}{\text{a}_0} = \Big(\frac{g_{\star,0}}{g_{\star, \rm rad}}\Big)^{1/3} \, \frac{T_{\rm CMB}}{T_{\rm rad}}\, e^{-N_I}\,, 
\end{align}
where $T_{\rm CMB} = 2.725$~K is the current-era Cosmic Microwave Background temperature, and $g_{\star, \,0})  \, (g_{\star, \, {\rm rad}})$ is the effective number of relativistic degrees of freedom today (during radiation), respectively. For the case of standard model, e.g., $g_{\star, 0} =3.36$~\cite{Planck}, and $g_{\star, {\rm rad}} =  106.75$. In general, for a temperature
$T$, $g_{\star}(T)$ evolves with time.  The radiation temperature $T_{\rm rad}$ that 
enters \eqref{scaling} is now given by \eqref{radreh}, which depends on the number of e-foldings of the Universe during the reheating period. This should be substituted 
in \eqref{scaling} in order determine $\text{a}_i$ to be used in our analysis of estimating the 
effects of the prolonged reheating period on PBHs densities, following \cite{kumar}.

We note though that the standard RVM prolonged reheating temperature 
\eqref{radmod} involves the dependence of the temperature on the ratio $\text{a}/\text{a}_{\rm eq}$. 
The formula was based on \eqref{evol}, which, as we explained above can only provide a scaling \eqref{scalreh} of the energy density during the reheating period $\rho_{\rm reh}=\rho_{\rm rad}$, as required for  
\eqref{scaling}, if its right-hand side is set to zero, that is, if the effects of the running vacuum in \eqref{evol} are not dominant. One may get an idea though on what values of 
$\text{a}/\text{a}_{\rm eq}$ correspond to the temperature \eqref{radreh}, in the case $w_{\rm reh}=1/3$, if 
we equate \eqref{radreh}  with the temperature 
given by \eqref{radmod}. This is done at the end of Appendix \ref{sec:appRVM}. 

However, in general, as already mentioned, 
one may have a different equation of state for the cosmic fluid during reheating, $w_{\rm reh}$ which could include that of matter. This, for instance, may happen in our stringy RVM model, if at the end of the RVM inflation, and the generation of chiral matter and radiation,
as explained above, one has approximately
\begin{align}\label{drho}
\dot \rho_{\rm RVM} \simeq 3\, w^\prime \, H \, \rho_f
\end{align}
with $\rho_f$ the energy density of the dominant reheating fluid, assumed a perfect one, appearing on the left-hand side of \eqref{evol}. The proportionality coefficient $w^\prime$ 
need not be negative, given that the standard decay of the running vacuum is replaced in the stringy RVM  by a potential increase as well, depending on the details of the generation of the chiral matter and radiation at the end of the inflationary period, which affect in general the sign of the coefficients \eqref{prolongcoeff}. Hence, for our purposes in this discussion 
we may assume $w^\prime$  to be a phenomenological real number.
In such a case, the RVM evolution-equation \eqref{evol} would become equivalent to that of the energy density of a perfect fluid 
\begin{align}\label{genericevol}
\dot \rho_f + 3(1 + w_f + w^\prime)\, H\, \rho_f \simeq 0\,,
\end{align}
The combination $w_f + w^\prime \equiv w_{\rm reh}$ could then be considered as the 
effective equation of state of the reheating fluid, $\rho_f=\rho_{\rm reh}$, 
in which case \eqref{genericevol} would correspond to the usual conservation equation of the stress tensor for the fluid.\footnote{The reader should notice, however, that to arrive at the scaling \eqref{scalreh}, which is required for the validity of  \eqref{radreh}, we need not actually identify the combination $w_{\rm reh}$ with a proper equation of state, in the sense of the corresponding pressure of the RVM vacuum during the generation of chiral matter and radiation, at the end of inflation, 
being characterised by $p^{\rm end}_{\rm RVM}= w^\prime \, \rho^{\rm end}_{\rm RVM}$. In fact, most likely this will not be the case, due to the complexity of the physical processes involved. Nonetheless, the mere validity of \eqref{drho}, leading to \eqref{genericevol}, suffices for studying the effects of the prolonged reheating on PBHs, as per the analysis of \cite{kumar} (and \cite{Cai}), which we review briefly in Appendix \ref{sec:appRVM}.}
We remark that, if we view $w_{\rm reh}$ as representing a perfect fluid equation of state, 
 then we may restrict ourselves to non negative $w_{\rm reh}$ for convenience, to avoid phantom energies. If in \eqref{genericevol} we consider that the initial $\rho_f$ was not a canonical radiation fluid, but one with an equation of state $w_f \ne 1/3$, we may end up 
with a $w_{\rm reh}=1/3$  for the effective cosmic fluid during the prolonged RVM reheating, and in general we may end up with various other possibilities for $w_{\rm reh}$, most importantly that of mimicking matter-dominance $w_{\rm reh}=0$.
It should be understood of course that eventually the intermediate reheating process characterised by \eqref{genericevol}, will be succeeded by ordinary radiation dominance,
and its duration should not be too long, lasting only for a few e-foldings $N_{\rm reh}$ until the complete exit from inflation. 

In this way, the generalised stringy RVM reheating period may admit the phenomenological description adopted in \cite{kumar} (and \cite{Cai}), which allows for estimates of the PBHs further enhancement as compared to the instantaneous reheating case. We review its features in Appendix \ref{sec:appRVM}.

\section{Conclusions \label{sec:concl}}

In this paper we have studied a model of inflation in the context of  Axion-Chern-Simons Gravity. In particular we have 
considered a two-axion string-inspired model, characterised by linear terms in their potential, but also containing non-perturbative
world-sheet-instanton-induced period terms, leading to modulation of the potential, as the time elapses. The linear terms might be due to condensation of
primordial GWs, which in turn lead to condensation of the gravitational Chern-Simons (anomalous) terms. In the context of the RVM inflationary model of \cite{ms}, it is the string-model independent KR axion that dominates in the early times (small scales), with the world-sheet instanton-induced potential terms of the other axion field arising from compactification eventually taking over, leading to fluctuations about the scale invariant spectrum during the inflationary era, and eventually ending inflation. For specific values of the parameters  we obtained an enhancement of the scalar power spectrum, 
as a consequence of the modulation of the potential due to the oscillatory terms arising as a result of the world-sheet instantons, which also results in an enhanced production of PBHs during inflation. The enhanced 
PBHs production results in the fact that in such models one can naturally sufficient abundance of PBHs (up to $70\%$) in the current era to account for a significant part of the DM in the Universe. 
In parallel, these models predicts GWs spectra with particular signatures, which could be falsified (or, hopefully, observed (!)) in  future 
experiments. 

Similar features charactrerise another class of two-stage inflationary models~\cite{silver}, based on string/brane-inspired axion monodromy models.
In the latter case, it is the compactification axion that drives the first inflationary phase, due to its linear potential, which is due to appropriate compactification (e.g. brane compactification in type IIB string models). This linear term dominates at the early stages of the first inflationary era, 
driven by this compactification axion field. 
Again, due to world-sheet instanton effects in such models, which dominate at later times,  there exists enhancement of PBHs and induced modifications of the profiles of GWs in the radiation era, of similar features as in the case of the models of \cite{ms}. The second stage of inflation comes when the KR axion field linear term in the potential (which might owe its existence to GW condensates as in the model of \cite{ms}) dominates. 
 In  these models, the mechanism leading to an enhanced power spectrum, and hence to enhanced densities of PBHs and GWs, is different from the one proposed in \cite{sasaki}, as here we consider it as a consequence of the inflection points~\cite{inflectionpoint,inflectionpoint2}  the pertinent axion potentials possess. 
Again in this case, PBHs can constitute a significant portion (up to $70\%$) of DM in such models.

As discussed above, there are, in principle, experimentally distinguishable differences between these two classes of inflationary models, as far as the morphology of the enhanced 
power spectra is concerned, which are imprinted in the energy densities of the corresponding GWs. In the first class of models~\cite{ms}, based on the hierarchy of scales \eqref{constr}, the intense oscillating patterns induced by the non-perturbatively-induced periodic terms in the axion potential, may lead to significant modifications of the profiles of the GWs  during the radiation era, which are in principle detectable at future interferometers. This is not the case for the smooth spectra of the second class of string-inspired axion-monodromy models~\cite{silver,sasaki}, based on the scale hierarchy \eqref{constr3}. Moreover, 
since the first class of models is an RVM cosmology~\cite{sola,lima}, and the RVM features survive, as explained in \cite{ms}, the exit from inflation and persist until the current era, one might obtain observable, in principle, deviations from $\Lambda$CDM in the modern epoch~\cite{rvmpheno,Tsiapi}, which may also alleviate the observed cosmological tensions, in particular the $H_0$ tension~\cite{rvmtens}, provided the latter do not admit mundane astrophysical or statistical explanations. These latter features do not seem to be shared by the stringy axion monodromy models. However, further detailed research along these lines needs to be performed before any definite conclusions are reached on these important matters. 

Finally we conclude by remarking that the above estimates of the PBH mass and fraction that plays the r\^ole of DM in the Universe are rather conservative, since we assumed that the reheating process of the string Universe was instantaneous. This is the case of conventional string axion monodromy scenarios~\cite{sasaki}, but may not always be the case for our stringy RVM model~\cite{ms}, examined in this work, where, for some choices of the microscopic parameters of the model, a prolonged reheating process due to the adiabatic decay of the RVM vacuum, which also produces most of the entropy of the Universe, takes place. In such a case, even if a matter-dominated reheating phase preceding the radiation era is absent, nonetheless the prolonged reheating process leading to radiation in the stringy RVM can affect the PBH mass function and therefore the fraction of PBHs that plays the r\^ole of DM in the Universe. In our discussion above we have assumed sets of parameters for which the reheating process is of very short duration (almost instantaneous), and from this point of view our estimates on the PBH densities are rather conservative.

\appendix
\section{Evaluation of fields perturbations}
\label{appendix1}
  
In order to obtain the scalar power spectrum we need to evaluate the linear perturbations. The equation for the perturbations of the fields $\phi_i+\delta \phi_i$ is given from the following equation~\cite{ringeval}:
\begin{equation}
\delta  \phi_i ^{\prime \prime}+(3-\epsilon) \delta  \phi_i ^\prime+\sum \frac{1}{H^2}\frac{\partial ^2 V}{\partial \phi_i \partial \phi_j}\delta \phi_i+\frac{k^2}{\text{a}^2H^2}\delta \phi_i=4 \Phi'\phi_i'- \frac{2\Phi}{H^2}\frac{\partial V}{\partial \phi_i}\, ,
\label{eq.pert}
\end{equation}
where $\text{a}$ is the scale factor of the expanding Friedman-Lema$\hat {\mathrm i}$tre-Robertson-Walker Universe, and $k$ denotes  the comoving wavenumber.  The Bardeen potential, $\Phi$, is given by the solution of equation
\begin{equation}
\Phi^ {\prime \prime}+(7-\epsilon)  \Phi^ \prime+ \left(    2 \frac{V}{H^2} + \frac{k^2}{\text{a}^2H^2} \right)\Phi=-\frac{1}{H^2}\frac{\partial V}{\partial \phi_i}.
\label{eq.Bardeen}
\end{equation}
The initial condition for the perturbation is given assuming that we are in Bunch-Davies vacuum
\begin{equation}
\text{a}\, Q \rightarrow \frac{e^{-ik\tau}}{\sqrt{2k}} \, , 
\end{equation}
where $Q$  is the Mukhanov-Sasaki variable \cite{ringeval,muksas}. The complete expression for the initial conditions for the perturbations,  as well as the initial conditions for the Bardeen potential are as 
\begin{equation}   
\begin{split}
\delta \phi_{i,ic}=\frac{1}{\sqrt{2 k}} \frac{1}{\text{a}_{ic}}, \quad \delta  \phi_{i,ic} '= -\frac{1}{\text{a}_{ic}\sqrt{2k}}\left(1+i\frac{k}{\text{a}_{ic}H_{ic}}\right)
\label{26b}
\end{split}
\end{equation}
and
\begin{equation}
\begin{split}
\Psi_{ic}=&\frac{1}{2 \left( {\varepsilon_H}_{ic} -\frac{k^2}{{\text{a}_{ic}}^2 H_{ic}^2} \right) } \left(   \phi_{i,ic}'  \delta {\phi_{i,ic}}'+\delta \phi_{i,ic}\left[3\left(\phi_{i,ic}'\right)^ +\frac{1}{ \text{a}_{ic}^2 H_{ic}^2}\left(\frac{dV}{d\phi_{i,ic}}\right)\right]\right)\\
 \Psi'_{ic}=& \frac{1}{2} \left( \phi_{i,ic}'\right) \delta \phi_{i,ic} -\Psi_{ic}\, ,
\end{split}
\end{equation}
where the suffix $ic$ denotes the initial condition at the beginning of inflation, as we defined it in the text.

After the evaluation of the linear perturbation, the power spectrum is given from the equation:
\begin{equation}
P_R=\frac{k^3}{(2 \pi)^2}|{\mathcal{R}_k}|^2
\end{equation}
where
\begin{equation}
\mathcal{R}_k=\Phi +   \sum_{\phi_i} \frac{\delta \phi_i }{\phi_i'}.
\end{equation}
are the scalar curvature perturbations.

\section{Early Stages of Cosmological Evolution of Running Vacuum Model }
\label{sec:appRVM} 

 In this section we elaborate further on reheating aspects of the RVM, which will help 
the reader clarify our conclusions in section \ref{sec:reh}.
The quantity $\text{a}_{\rm eq}$ appearing in \eqref{radmod} is determined by the boundary conditions of the problem at the onset of the RVM inflation. Indeed the modified evolution of an RVM cosmic fluid (``$f$''), which in addition to the vacuum contributions to the energy density (``RVM'') is also characterised by an excitation perfect fluid (``m'') with equation of state $w_m$, reads (the overdot denotes cosmic-time derivative)~\cite{lima}:
\begin{align}\label{evol}
\dot \rho_f  + 3(1+ w_f)\, H\, \rho_f = - \dot \rho_{\rm RVM}\,, \qquad  
\rho_{\rm RVM} = 3 \kappa^{-2} \Big(\nu\, H^2 +  \alpha \, \frac{H^4}{H_I^2}\Big)\,, 
\end{align}
where $\nu$, $\alpha$ are constants ({\it cf.} \eqref{rvmener}), and the reader should notice that in the stringy RVM case $c_0=0$, sine a cosmological constant is incompatible with perturbative and non-perturbative (swampland~\cite{swamp}) string theory. Only metastable de Sitter vacua are allowed, in agreement with the stringy RVM nature.
Eq. \eqref{evol} is an energy-density conservation equation, taking into account the 
contributions from the running vacuum, which change with the cosmic time.

When combined with the equations of motion, Eq. \eqref{evol}  leads to the RVM evolution equation for the Hubble parameter:
\begin{align}\label{evolHubble}
\dot H + \frac{3}{2} \, (1 + w_f) \, H^2 \, \Big[ 1 - \nu - \alpha \, \frac{H^2}{H_I^2}\Big] =0\,,
\end{align}
which admits as a solution:
\begin{align}\label{evolHsol}
H (\text{a}) = \Big(\frac{1-\nu}{\alpha}\Big)^{1/2} \, \frac{H_I}{\sqrt{\mathcal D \, a^{3(1 + w_f)} + 1}}\,,
\end{align}
with $\text{a}$ the scale factor of the universe in units of today's scale factor $\text{a}_0$, and 
$\mathcal D > 0$ is an integration constant, which in terms of the microscopic stringy RVM depends on the boundary conditions at the onset of the RVM inflation~\cite{ms}, and in the low-energy effective theory is viewed as a phenomenological parameter. 
In the stringy RVM we have ({\it cf.} \eqref{rvmener} ): $|\nu| \sim 10^{-3}$ and $\alpha = \mathcal O(1)$~\cite{ms}, 
which we can use from now on. 

From \eqref{evolHsol} we indeed observe that for early times $\mathcal D \, \text{a} \ll 1$ one obtains a metastable de Sitter spacetime, with approximately constant $H$.
The analysis in \cite{lima} has shown that the vacuum energy density and the excitation fluid energy density $\rho_f$ are given by:
\begin{align}\label{energydens}
\rho_{\rm RVM} &= \frac{\rho_I}{\Big(\mathcal D \, \text{a}^{3(1+w_f)} + 1\Big)^2}\,,  \qquad 
\rho_f= \frac{\rho_I\, \mathcal D \, \text{a}^{3(1+w_f)}}{\Big(\mathcal D \, \text{a}^{3(1+w_f)} + 1\Big)^2}\,, \qquad
\rho_I \equiv 3\, \kappa^{-2} \, H^2\,.
& \Rightarrow 
\end{align}
We have therefore
\begin{align}\label{ratio1}
\frac{\rho_{\rm RVM}}{\rho_f} =  \mathcal D^{-1}\, \text{a}^{-3(1+w_f)}\,.
\end{align}
In the case of radiation ``$f$''$=\rm rad$ (as is relevant for RVM)~\cite{lima}, including the stringy one~\cite{ms}, $w_f=1/3$, hence one obtains for the ratio of vacuum to radiation energy densities:
\begin{align}\label{ratio2}
\frac{\rho_{\rm RVM}}{\rho_{\rm rad}} = \mathcal D^{-1} \, \text{a}^{-4}\, ,
\end{align}
which implies that at the equivalence point between vacuum and radiation $\text{a}_{\rm eq}$
one has 
\begin{align}\label{Daeq}
\mathcal D \, \text{a}^4_{\rm eq} =1\,, 
\end{align}
and therefore \eqref{evolHsol} can be expressed in terms of the parameter $\text{a}_{\rm eq} $ as:
\begin{align}\label{evolHsol2}
H (\text{a}) = \frac{H_I}{\sqrt{\Big(\frac{\text{a}}{\text{a}_{\rm eq}}\Big)^4 + 1}}\,.
\end{align}
This expression, together with the corresponding expression for the radiation density
\begin{align}\label{rad}
\rho_{\rm rad} = \rho_I \, \frac{(\text{a}/\text{a}_{\rm eq})^4}{\Big[1 + (\text{a}/\text{a}_{\rm eq})^4 \Big]^2} =
\frac{\pi^2}{30}\, g_\star \, T_{\rm rad}^4\,,
\end{align} 
is used~\cite{lima} to arrive at the Equation \eqref{radmod} in the main text, where $g_\star$ is the effective number of relativistic degrees of freedom at radiation era (in general~\cite{kolb}, $g_\star (T)$ is a function of the cosmic temperature). 

Some discussion is in due here, concerning the prolonged reheating period in the RVM framework, including the stringy RVM,  prior to the radiation era, as compared to generic, non-RVM cosmologies.

In the phenomenological approach of \cite{kumar,Cai}, the reheating process is assumed to be dominated by some perfect fluid with an effective equation of state $w_{\rm reh}$, which may assume other values than that of radiation. Let us see whether this can be realised in the RVM framework.

During reheating in the standard RVM framework~\cite{lima}, 
that is the linear increasing phase of the scale factor with the temperature 
in \eqref{radmod}, for $\text{a} < \text{a}_{\rm eq}$, we may assume the evolution equation  
of the form \eqref{evol}, with $f$ being given by a radiation fluid. 
From \eqref{energydens}, \eqref{ratio1} (or \eqref{ratio2} for radiation fluid component) and \eqref{evol}, we then obtain :
\begin{align}\label{modevol}
\dot \rho_f + 3\, H\, (1 + w_f) \, \Big[1 - \frac{2}{(\text{a}/\text{a}_{\rm eq})^{3(1+w_f)} + 1}\Big]\, \rho_f =0 \,,
\end{align}
where we used the fact that the equivalence point between the fluid $f$ (radiation in conventional RVM) and vacuum, that is the equality of the corresponding vacuum energy densities, implies 
$\mathcal D \, \text{a}_{\rm eq}^{3(1+w_f)} =1$. 
We observe from \eqref{modevol} that the expectation of \cite{kumar,Cai} for a (prolonged) reheating phase being described by a perfect fluid 
 with an effective equation of state
$w_{\rm reh}$, and energy density scaling as:
\begin{align}\label{scalreh}
\rho_{\rm reh} \sim \text{a}_{\rm reh}^{-3(1 + w_{\rm reh})} \,,
\end{align}
is not met in the standard RVM , due to complicated scaling with $a$, in the reheat regime 
$\text{a} < \text{a}_{\rm eg}$ ({\it cf.} fig~\ref{fig:Tvsa} and Eq.~\eqref{radmod}). 

However, in our stringy RVM, we have the axion matter (both KR and compactification axions) which may play a r\^ole in the reheating, together with the cancellation of the primordial gravitational anomalies by the gravitational anomalies of 
the chiral matter generated during the end of inflation~\cite{ms}. As discussed in the text, the almost instantaneous decay of the compactification axion into radiation, in our set of parameters ({\it cf.} Table \ref{table1}) will induce a quick reheating, shortening significantly the duration of the epoch in which the scale factor increases linearly with 
the temperature \eqref{radmod}, thus having $\text{a}_{\rm eq} \ll \text{a} \ll 1$, implying that the approximation of the instantaneous reheating is a good one for this set, which lead to the estimates for the enhanced scalar power spectrum and the density of PBH presented in section \ref{sec:2mon}. 

On the other hand, given that the scale $\Lambda_1$ in the effective axion potential ({\it cf.} \eqref{effpot}) depends on microscopic properties of the underlying string model, such as the world-sheet instanton actions, one may consider other possibilities with a prolonged reheating period, in which the axion decay to radiation is delayed significantly, and the reheating period is dominated by matter, with an equation of state $w_f=0$. 

In general, as discussed in section \ref{sec:reh}, we may assume that in some regimes of the parameters (string landscape), in our model, the RVM evolution \eqref{evol}
during the reheating period could be represented by that of  
 an effective fluid ``$f$''$=\rm  reh$, Eq.~\eqref{genericevol}. 
 This would
then make plausible the assumption that during reheating, one would have a (perfect) fluid with an effective equation of state
$w_{\rm reh}$, and energy density scaling as \eqref{scalreh}.
From \eqref{reheat} in section \ref{sec:reh}, then, we may write 
\begin{align}\label{scaling2}
\frac{\rho_{\rm reh}}{\rho_{\rm end}} = \Big(\frac{\text{a}_{\rm reh}}{\text{a}_{\rm end}}\Big)^{-3(1 + w_{\rm reh})} = e^{-3\, (1 + w_{\rm reh})\, N_{\rm reh}}\,,
\end{align}
with  $N_{\rm reh}$ the number of e-foldings during the part of the RVM inflation in which reheating occurs, and $\text{a}_{\rm end}$ denoting the end of inflation before reheating.
In the stringy RVM model, this reheating phase is characterised by the cancellation of primordial gravitational anomalies by the ones generated by chiral matter~\cite{ms}. As argued in section \ref{sec:reh}, this is consistent with the suppression  of $\dot \rho_{\rm RVM}$ terms in the respective evolution equation \eqref{evol} with the replacement of ``$f$'' by the effective reheating fluid ``$\rm reh$''. 

In such phenomenological situations, one may then identify $\rho_{\rm reh} = \rho_{\rm rad}$, with $\rho_{\rm rad}$ given by \eqref{rad}, to determine, by means of the Friedman equation
for the energy density at the end of inflation, $\rho_{\rm end} = 3\, \kappa^{-2}\, H_{\rm end}^2$, the temperature at the onset of radiation in terms of the reheat parameters $N_{\rm reh}, w_{\rm reh}$~\cite{kumar} 
\begin{align}\label{radreh}
T_{\rm rad} = \Big(\frac{90\, H_{\rm end}^2 \, \kappa^{-2}}{\pi^2\, g_{\star, \rm rad}} \, \Big)^{1/4}\,
\Big(e^{-3\, (1 + w_{\rm reh})\, N_{\rm reh}}\, \Big)^{1/4}\,,
\end{align}
where we used \eqref{scaling2}. This can be used in generic models to obtain the initial value of the scale factor $\text{a}_i$ that is used in the determination of the scalar power spectra
and the densities of PBHs, as discussed in \cite{kumar} and reviewed briefly in
section \ref{sec:reh}.

Let  us discuss below under which circumstances such an effective description 
is appropriate within the RVM framework. 
Assuming that during the RVM inflation $H$ remains approximately constant, one may replace $H_{\rm end} \simeq H_I$, the inflationary 
de Sitter scale~\cite{Planck}. 

As discussed in the main text, in section \ref{sec:reh}, using \eqref{radmod} 
at the onset of the radiation period, that is identifying $T_{\rm rad}$ in \eqref{radreh} with the right-hand side of \eqref{radmod}, 
we can determine the values of the ratio $\text{a}_{\rm rad}/\text{a}_{\rm eq}$ (in terms of the reheating parameters
$w_{\rm reh} = 1/3, N_{\rm reh} > 0$) that would correspond formally to the temperature \eqref{radreh}. 
Such an identification is only formal, because as we have explained above, 
the conventional RVM reheating leading to \eqref{radmod} may not be valid in string theory, and moreover does not correspond to a perfect-fluid reheating phase. Nonetheless, the identification would yield some information as to how far from the predictions of the standard RVM we would be by assuming the perfect reheating scaling \eqref{scalreh}. 
This identification of the radiation temperatures then 
yields the 
algebraic equation
\begin{align}\label{eqalg}
x^4 - e^{-3(1+w_{\rm reh})\, N_{\rm reh}} \Big( 1 + x^4\Big)^2 =0, \qquad x \equiv \frac{\text{a}_{\rm rad}}{\text{a}_{\rm eq}} > 0\,, \quad w_{\rm reh} = 1/3\,.
\end{align}
Observing that 
\begin{align}\label{exponent}
\exp(b^2) = \exp(3\, [1+w_{\rm reh}]\,N_{\rm reh} ) = \exp( 4 \, N_{\rm reh}) \gg 1\,, 
\end{align}
already for a few $N_{\rm rreh}$ e-foldings, 
we have the potentially real solutions of \eqref{eqalg}, as given by Mathematica:
\begin{align}\label{soln}
x_1 &=  - \Big(-1 + \frac{e^{b^2}}{2}\, (1 + \sqrt{1-4\, e^{-b^2}}\Big)^{1/4}\, , \nonumber \\
x_2 &=   + \Big(-1 + \frac{e^{b^2}}{2}\, (1 + \sqrt{1-4\, e^{-b^2}}\Big)^{1/4}\, , \nonumber \\
x_3 &=  -  \Big(- 1 + \frac{e^{b^2}}{2}\, (1 - \sqrt{1 - 4\, e^{-b^2}} \Big)^{1/4}\, , \nonumber \\
x_4 &=  + \Big(- 1 + \frac{e^{b^2}}{2}\, (1 - \sqrt{1 - 4 \, e^{-b^2})} \Big)^{1/4}\,.
\end{align}
with the reality condition that $4e^{-b^2}  < 1$ amply satisfied.\footnote{Out of completion in our discussion we would like to point out that these equations do not admit a solution for $b^2=0$ which is the RVM de Sitter equation of state. This stems from the fact that in such a fluid the vacuum completely dominates and there is no flow of $H$ to allow for 
the presence of radiation component (or in fact any other equation of state component in \eqref{evol}), whose only solution would be the standard de Sitter space.} 
On account of \eqref{exponent}, then, we observe that 
there are two positive $x>0$ solutions\footnote{That a fixed temperature value in \eqref{radmod} corresponds to two values of the ratio $a/a_{\rm eq}$ is due to the shape of the $T-a$ curve, see figure \ref{fig:Tvsa}.} for a few $N_{\rm reh} >0$ in the range $N_{\rm reh} \in [5, 10]$, as required in the analysis of \cite{kumar} to get significant increase in the PBH densities and mass functions. These are the $x_{2,4}$ solutions of \eqref{soln}, which, after an appropriate  large-$b^2 \gg 1$ expansion of the square root, yield: 
\begin{align}\label{soln2}
x_2 = \frac{\text{a}_{\rm rad}}{\text{a}_{\rm eq}}{\Big |_2} 
\simeq e^{b^2/4} = e^{N_{\rm reh}} \, \qquad 
x_4  = \frac{\text{a}_{\rm rad}}{\text{a}_{\rm eq}}{\Big |_4} \simeq e^{-b^2/4} = e^{- N_{\rm reh}} = x_2^{-1} \,. 
\end{align}
Physically, we want the smallest value of the ratio $\frac{\text{a}_{\rm rad}}{\text{a}_{\rm eq}} <1 $ to describe the onset of the radiation-dominating-reheating phase, as  this is the reheating phase of the RVM vacuum and in this case the RVM vacuum still dominates over the fluid (radiation) component, and thus one can still have an exponential growth of the scale factor. The closest the value of $\text{a}/\text{a}_{\rm eq}$ to 1, the better the agreement of the predictions of the RVM with the single-perfect-fluid reheating model.
To get an idea of the numbers involved,  we discuss below two  indicative cases. For the case $N_{\rm reh}=5$, for instance, we obtain:
$\frac{\text{a}_{\rm rad}}{\text{a}_{\rm eq}}{\Big |_4} =  6.7 \times 10^{-3} \quad (\rm with ~\frac{\text{a}_{\rm rad}}{\text{a}_{\rm eq}}{\Big |_2} = 148.4))$,
while for $N_{\rm reh}=10$ we have:
$\frac{a_{\rm rad}}{a_{\rm eq}}{\Big |_4} =  4.5 \times 10^{-5}, \quad (\rm with~ \frac{a_{\rm rad}}{a_{\rm eq}}{\Big |_2} = 2.2 \times 10^4)$. 
This completes our discussion.

\acknowledgments

We thank J. Fumagalli for a useful correspondence. 
The work of N.E.M. is supported in part by 
the UK Science and Technology Facilities  research Council (STFC) under the research grant ST/T000759/1. 
That of V.C.S. and I.D.S is supported by the Hellenic Foundation for Research and Innovation (H.F.R.I.)
under the “First Call for H.F.R.I. Research Projects to support Faculty members and Re-
searchers and the procurement of high-cost research equipment grant” (Project Number:
824). N.E.M.  also acknowledges participation in the COST Association Action CA18108 ``{\it Quantum Gravity Phenomenology in the Multimessenger Approach (QG-MM)}''.

\end{document}